\documentclass[journal]{IEEEtran}
\IEEEoverridecommandlockouts
\usepackage{mathrsfs}
\usepackage{amsfonts}
\usepackage{ifpdf}
\usepackage{cite}
\ifCLASSINFOpdf
\usepackage[pdftex]{graphicx}
\else \fi
\usepackage{amsmath}
\usepackage{array}
\usepackage{mdwmath}
\usepackage{mdwtab}
\usepackage{eqparbox}
\usepackage[tight,footnotesize]{subfigure}
\usepackage{algorithmic}
\usepackage{algorithm}
\usepackage{amsmath}
\usepackage{epsfig}
\usepackage{ae}
\usepackage{amssymb}
\usepackage{times}
\usepackage{amsmath}   
\usepackage{booktabs}  
\usepackage{threeparttable} 
\usepackage{multirow}       
\usepackage{xcolor}

\usepackage{amsmath,amssymb,amsfonts}
\usepackage{graphicx}
\usepackage{textcomp}
\usepackage{xcolor}

\newtheorem{definition}{\textbf{Definition}}
\newtheorem{remark}{\textbf{Remark}}

\def\BibTeX{{\rm B\kern-.05em{\sc i\kern-.025em b}\kern-.08em
    T\kern-.1667em\lower.7ex\hbox{E}\kern-.125emX}}

\begin{document}

\title{Optimal Status Update for Caching Enabled IoT Networks: A Dueling Deep R-Network Approach}


%
%

%
\author{
Chao Xu, Yiping Xie, Xijun Wang, \\ Howard H. Yang, Dusit Niyato, and Tony Q. S. Quek


\vspace{-0.2cm}
\thanks{

C. Xu and Y. Xie are with School of Information Engineering, NWAFU, Shaanxi, China (e-mail: cxu@nwafu.edu.cn).

X. Wang is with School of Electronics and Information Technology, SYSU, Guangzhou, China (e-mail: \mbox{wangxijun@mail.sysu.edu.cn}).

H. Yang is with the Zhejiang University/UIUC Institute, Zhejiang University, Haining, China, and was with the ISTD Pillar, SUTD, Singapore (e-mail: eehowardh@gmail.com).

D. Niyato is with the School of Computer Science and Engineering, NTU, Singapore (e-mail: \mbox{DNIYATO@ntu.edu.sg}).

T. Q. S. Quek is with the ISTD Pillar, SUTD, Singapore (e-mail: tonyquek@sutd.edu.sg).

}
}

\maketitle
\begin{abstract}
In the Internet of Things (IoT) networks, caching is a promising technique to alleviate energy consumption of sensors by responding to users' data requests with the data packets cached in the edge caching node (ECN). However, without an efficient status update strategy, the information obtained by users may be stale, which in return would inevitably deteriorate the accuracy and reliability of derived decisions for real-time applications. In this paper, we focus on striking the balance between the information freshness, in terms of age of information (AoI), experienced by users and energy consumed by sensors, by appropriately activating sensors to update their current status. Particularly, we first depict the evolutions of the AoI with each sensor from different users' perspective with time steps of non-uniform duration, which are determined by both the users' data requests and the ECN's status update decision. Then, we formulate a non-uniform time step based dynamic status update optimization problem to minimize the long-term average cost, jointly considering the average AoI and energy consumption. To this end, a Markov Decision Process is formulated and further, a dueling deep R-network based dynamic status update algorithm is devised by combining dueling deep Q-network and tabular R-learning, with which challenges from the curse of dimensionality and unknown of the environmental dynamics can be addressed. Finally, extensive simulations are conducted to validate the effectiveness of our proposed algorithm by comparing it with five baseline deep reinforcement learning algorithms and policies.
\end{abstract}
\begin{IEEEkeywords}
Internet of Things, age of information, deep reinforcement learning, dynamic status update, non-uniform time step, long-term average reward.
\end{IEEEkeywords}

\section{Introduction}

Acting as a critical and integrated infrastructure, the Internet of Things (IoT) enables ubiquitous connections for billions of things in our physical world, ranging from tiny, resource-constrained sensors to more powerful smart phones and networked vehicles\cite{Survey_IoT_Applications_2015}. In general, the sensors are powered by batteries with limited capacities rather than fixed power supplies. Thus, to exploit the benefits promised by IoT networks, it is essential to address the energy consumption issue faced by sensors. Recently, caching has been proposed as a promising solution to lower the energy consumption of sensors by reducing the frequency of environmental sensing and data transmission \cite{IoT_Caching_2016_Network,Caching_IoT_EH_ICC_2016,IoT_Caching_2017_RL}. Particularly, by storing the data packets generated by sensors at the edge caching node (ECN), e.g., access point (AP) or mobile edge node, users could retrieve the cached data directly from the ECN instead of activating sensors for status sensing and data transmission, thereby lowering their energy consumption.

Caching multimedia contents at the edge of wireless networks (e.g., the ECN or user equipment) has been recognized as one of the promising technologies for the fifth Generation (5G) wireless networks and hence, well studied in existing work \cite{Femtocell_Caching_2013,Content_Caching_2014,Our_Caching_Comm_Mag,Caching_Multicasting_BZhou_2016,Tang_2019,Update_WN_Lifetime_JSAC_2018}.
However, compared with the multimedia contents (e.g., music, video, etc.) in traditional wireless networks, the data packets in IoT networks have two distinct features: 1) The sizes of data packets created by IoT applications are generally much smaller than those of multimedia contents. Therefore, for IoT networks, the storage capacity of each ECN is generally sufficient to store the latest status updates generated by all sensors. 2) For many real-time IoT applications, the staleness of information at the user side can radically deteriorate the accuracy and reliability of derived decisions. Such concerns arise from a variety of real-time IoT applications. For instance, in an environmental monitoring system, battery powered sensors are often adopted for data collection, including the temperature, humidity, $CO_2$ concentration, or illumination intensity. The collected data can be cached in the ECN and consumed by users for further computation and analysis. Also, in an intelligent transportation system, information about the level of traffic congestion or availability of parking lots can be obtained by sensors, and then cached in the roadside unit (RSU) to reduce the sensors' energy consumption. In this case, the information freshness is of utmost importance for nearby drivers.

To this end, the main concern for edge caching enabled IoT networks would be related to how to properly update the cached data to \textit{lower the energy consumption} of sensors and meanwhile \textit{improve the information freshness} at users or, in other words, to refresh the cached items in an energy-efficient and timely fashion. We note that the previous work \cite{Update_WN_Lifetime_JSAC_2018} introduced the concept of lifetime to depict the time popularity variation of social media contents, and investigated the content caching strategy for a single user equipment to minimize the transmission cost.
However, in \cite{Update_WN_Lifetime_JSAC_2018} all unexpired contents were equally treated, regardless of their staleness. Hence, the devised status update policy does not apply to caching enabled IoT networks carrying real-time applications, for which the fresher data is potentially more informative.

Recently, Age of Information (AoI) has been proposed as a metric to assess the information freshness by measuring the time elapsed since the latest received packet was generated from the source\cite{AoI_Org_2012,AoI_Survey_2017,YanArafaQue:20}, which, as presented in studies \cite{Our_IF_2019,BZhou_Samp_Up_2019,BZhou_Non_Uni_Pack_2020}, could be optimized by adequately adjusting the status update procedure of sensors. Particularly, authors in \cite{Our_IF_2019} considered an edge computing enabled IoT network, and developed a generating set search based algorithm to optimize sensors' update rates to minimize the achieved maximum average peak AoI. In \cite{BZhou_Samp_Up_2019}, the sampling and transmission processes are jointly optimized to minimize the average AoI at the destination under the average energy constraints for individual sensors. Also focusing on the joint sampling and transmission policy design for IoT networks, work \cite{BZhou_Non_Uni_Pack_2020} extended \cite{BZhou_Samp_Up_2019} by considering the different sizes of packets generated by different sensors.

Armed with this metric, a few recent efforts\cite{Age_Updating_2017,Refresh_Rate_AoI_2018,AoI_Cache_Updating_2019,Ave_AoI_EH_Sensor_Pappas} have developed strategies that improve data freshness in caching enabled IoT networks by minimizing the AoI. Specifically, the authors in \cite{Age_Updating_2017} proposed a status update algorithm to minimize the popularity-weighted average of AoI values at the cache. Then, in \cite{Refresh_Rate_AoI_2018}, the optimal data refresh rate allocation policy was derived, with which the popularity-weighted average AoI at the cache was minimized. Study \cite{AoI_Cache_Updating_2019} further extended \cite{Age_Updating_2017} by considering the relation between the time consumption for one update and the corresponding update interval. Finally, in \cite{Ave_AoI_EH_Sensor_Pappas}, the authors considered a caching enabled IoT system with one energy harvesting sensor, and investigated the average AoI for a probabilistic update policy at the ECN.
However, in \cite{Age_Updating_2017,Refresh_Rate_AoI_2018,AoI_Cache_Updating_2019,Ave_AoI_EH_Sensor_Pappas}, the AoI was evaluated from the ECN's perspective instead of from the perspective of individual users. In fact, it is more essential to optimize the AoI of users, since they are the real data consumers and the ultimate decision makers.

Recognizing this, some recent studies begin to explore schemes to improve the information freshness experienced by users in caching enabled IoT networks\cite{ILP_Caching_ICC_2020,IF_Caching_2020,AoI_Dealy_Tradeoff_2020,DRL_Updating_2020,IoT_Caching_2020}. In \cite{ILP_Caching_ICC_2020}, to address the trade-off between the status update and AoI costs, the caching placement problem was formulated as a static integer linear program, where the arriving time of users' requests was assumed to be known non-causally to the ECN. For cached items, the information freshness was maximized by jointly optimizing the update rate of the ECN and request frequency of the user in \cite{IF_Caching_2020}. In \cite{AoI_Dealy_Tradeoff_2020}, a threshold based data update policy was devised to serve users on demand. Assuming that users' requests to be served in the future were known in advance, authors in \cite{DRL_Updating_2020} proposed a dynamic content update algorithm to minimize the average AoI of data delivered to users plus the cost related to the content updating. In \cite{IoT_Caching_2020}, by modeling users' requests with the time-varying content popularity at the ECN, a dynamic status update policy was derived for ECNs under the coordination of the cloud, where all sensors were allowed to transmit their generated status updates simultaneously without collisions.

In this work, we develop a dynamic status update strategy to balance the AoI experienced by users and energy consumed by sensors in caching enabled IoT networks, where neither the request patterns of users nor the transmission failure probabilities of sensors are known to the ECN. Particularly, we consider a time-slotted IoT network consisting of an ECN, multiple sensors, and multiple users, where the ECN would store the latest data packets generated by sensors. The decision epochs, also referred to as time steps\cite{Non_Uniform_Time_Step_2020}, are of non-uniform duration, i.e., the duration of one time step is determined by the data requests of users and the status update decision made by the ECN.
To strike a balance between the average AoI experienced by the users and energy consumed by the sensors, we formulate \textit{a non-uniform  time step} based dynamic status update optimization problem to minimize the expectation of the\textit{ long-term average cost}, by jointly accounting for the average AoI and energy consumption. To solve this problem, we cast the corresponding status update procedure as a Markov Decision Process (MDP), where the number of state-action pairs increases exponentially with respect to the number of sensors and users. Consequently, we concentrate on the deep reinforcement learning (DRL) algorithm design, with which the curse of dimensionality can be circumvented. Specifically, we have developed a dueling deep R-network based dynamic status update (DDR-DSU) algorithm
to solve the originally formulated problem, which combines the dueling deep Q-network (DQN) \cite{Dueling_DQN_Org} and R-learning \cite{RL_1993}.
As a tabular model-free reinforcement learning (RL) algorithm, R-leaning was tailored for maximizing the long-term undiscounted average reward for small-scale decision-making problems.
To this end, our proposed DDR-DSU algorithm requires no prior knowledge of the dynamics of the MDP and can be used to directly maximize the long-term average reward in lieu of the discounted one, whose effectiveness is further verified via extensive simulation results.

It is worth noting that, in theory, maximizing the discounted long-term cumulative reward is not the same as maximizing the long-term average reward for continuing MDPs (i.e.,  without termination states), unless in the limiting regime where the discount factor $\gamma$ approaches 1. As such, it is extremely challenging to use existing DRL algorithms, initially developed to optimize the discounted cumulative reward, to maximize the average reward, since the discounted reward would go to infinity as $\gamma$ approaches 1. In fact, how to design DRL algorithms that maximize the long-term average reward is still an open problem. Interested readers are suggested to see our technical report \cite{Our_TR_2021} for detailed proofs and discussions. To the best of our knowledge, this is the first work that develops DRL algorithms to maximize the long-term average reward for continuing MDPs by integrating R-learning with traditional DRL algorithms.

While our conference version \cite{CXu_Updating_2020} also dealt with the same problem, it had three limitations: 1) The concerned problem scale was small and hence, was addressed with a traditional tabular model-free RL algorithm, i.e., expected Sarsa. 2) The duration of different time steps was assumed to be uniform, i.e., irrespective of the users' requests and selected actions. 3) The aim was to maximize the discounted long-term cumulative reward, which is not effective enough for improving the real system performance, as demonstrated by simulation results in Section \ref{Sec:Simulation}. This work removes several assumptions in \cite{CXu_Updating_2020}, and the main contributions can be summarized as follows.
\begin{itemize} \setcounter{enumi}{0}
\item We formulate the concerned dynamic status update procedure in caching enabled IoT networks as an MDP with non-uniform time steps, where the dynamics of the environment is unknown. Within each time step, we depict the evolution of the AoI value with each sensor and user pair for different time slots to well amortize the AoI related cost over the corresponding time duration. Furthermore, the AoI related cost for each individual user is derived for non-uniform time steps.
\item We develop a dueling deep R-network based dynamic learning  algorithm, termed DDR-DSU, to directly maximize the obtained long-term average reward, by combining the dueling DQN and R-learning. Using the proposed framework, R-learning could be also combined with various traditional DRL algorithms, e.g., deep R-network (DRN) can be devised by combining R-learning with the standard DQN. Such DRL algorithms modified by incorporating R-learning can also be implemented for other decision-making tasks to maximize \textit{the long-term average reward}. In contrast, traditional DRL algorithms generally are devised to optimize the \textit{discounted long-term cumulative reward}.
\item The effectiveness of our proposed DDR-DSU algorithm is verified with extensive simulation results, where a few baseline DRL algorithms and policies are compared, i.e., the deep R-network based dynamic status update (DR-DSU), dueling DQN based dynamic status update (DDQ-DSU), DQN based dynamic status update (DQ-DSU), greedy policy, and random policy. It is interesting but not surprising to see that, for DQN based algorithms, i.e., DDQ-DSU and DQ-DSU, the performance in terms of both the convergence rate and achieved average reward is dramatically affected by changing the adopted discount factor, a hyperparameter. In contrast, for their counterparts modified by incorporating R-learning, i.e., DDR-DSU and DR-DSU, both the stability and performance can be significantly improved in terms of the achieved average reward. Besides, incorporating the dueling network architecture into DR-DSU can further improve the convergence rate of the proposed DDR-DSU algorithm.
\end{itemize}

The organization of this paper is as follows. In Section II, the description of the system model and the formulation of the dynamic status update optimization problem are presented. In Section III, we cast the dynamic status update procedure as an MDP and then, develop a dueling deep R-network based algorithm to solve it. Extensive simulation results are presented to show the effectiveness of our proposed scheme in Section IV, and conclusions are drawn in Section V.

\section{System model and problem formulation} \label{Sec:Section 2}

\subsection{Network model}

\begin{figure} [!t]
\centering
\subfigure[] {\leavevmode \epsfxsize=3.0in  \epsfbox{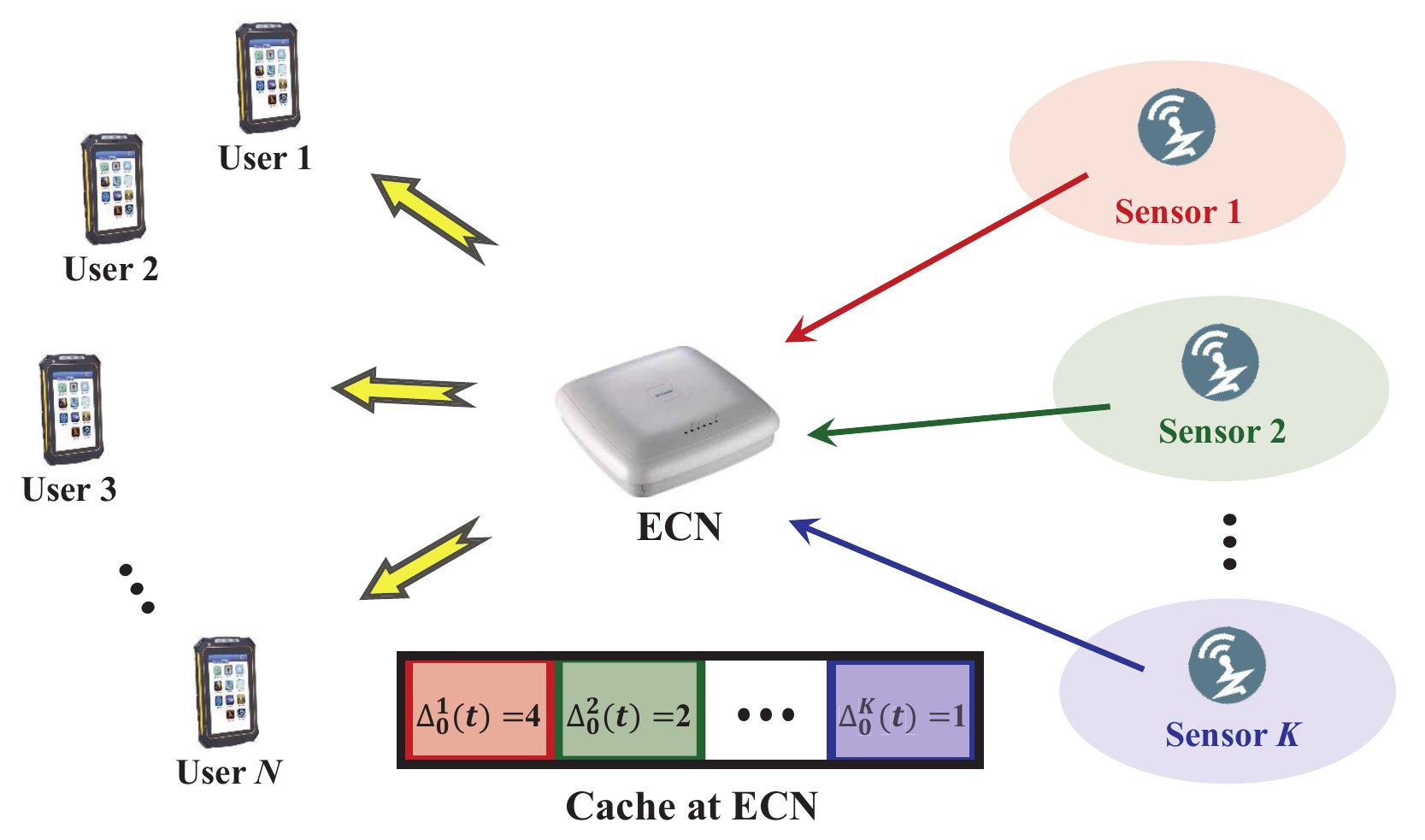}}
\subfigure[] {\leavevmode \epsfxsize=3.0in  \epsfbox{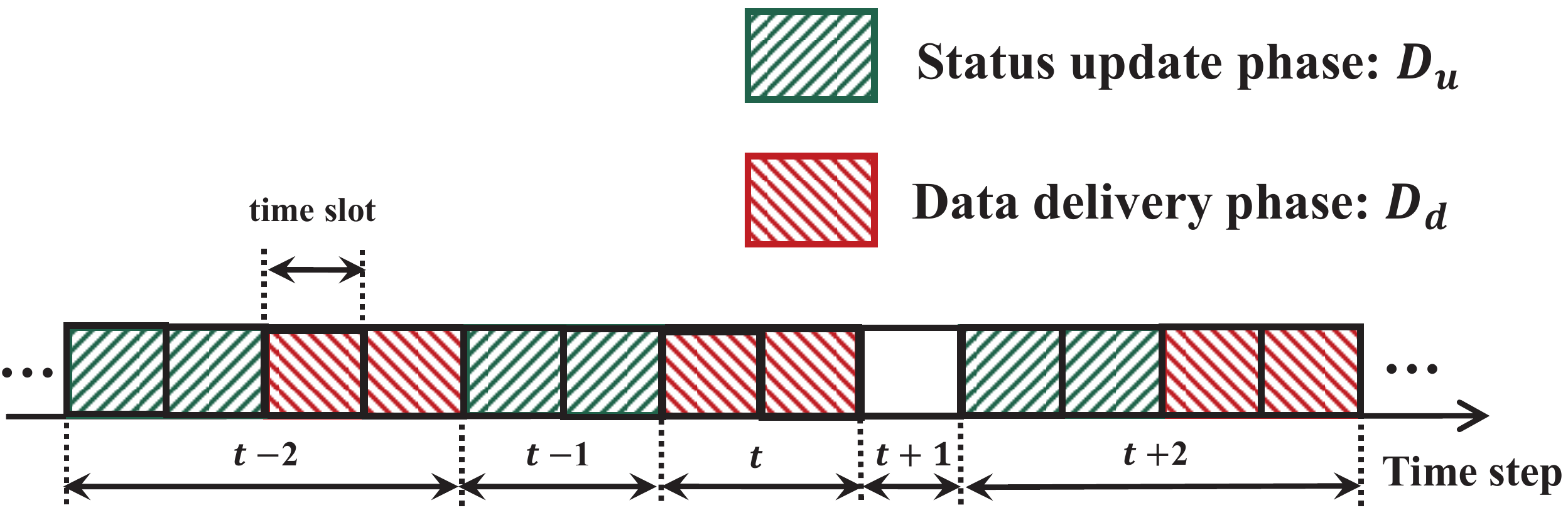}}
\centering \caption{(a) An illustration of a caching enabled IoT network, where $\Delta_0^k(t)$ denotes the AoI of sensor $k$ at the ECN at the beginning of each time step $t$. (b) An illustration of the structure of non-uniform time steps, where SUP and DDP consist of $D_u$ and $D_d$ time slots, respectively.} \label{Fig:System_Model_Time_Steps}
\end{figure}

We consider an IoT network consisting of one ECN, $N$ users, and $K$ sensors, as illustrated in Fig. \ref{Fig:System_Model_Time_Steps} (a). This model arises from applications such as the real-time environmental monitoring system for smart agriculture or the real-time traffic monitoring system for intelligent transportation.
The user and sensor sets are respectively denoted by $\mathcal N = \{1, 2, \ldots, N\}$ and $\mathcal K = \{1, 2, \ldots, K\}$.
We consider a discrete-time system, in which the time axis is segmented into slots with equal length. In this IoT network, the ECN erratically asks the sensors to make status updates during the status update phase (SUP), and on the other hand, responds to the users' demands, if any, with the cached packets during the data delivery phase (DDP). The time duration of SUP and DDP are assumed to be fixed as $D_u$ and $D_d$ slots, respectively. Without loss of generality, we consider that $D_u$ and $D_d$ are two integers.\footnote{It is worth noting that although the time durations of SUP and DDP are fixed, our system model still captures the heterogeneity of users and sensors. Particularly, for users, their data request probabilities are allowed to be different. Meanwhile, as presented in  (\ref{Eq:AoI_Cost_Per_Slot}), the weights for evaluating their impacts on the AoI cost can also be different. Additionally, for sensors, the heterogeneity can be characterized by the content popularity, transmission failure probability, and energy consumption for completing one status update.} Accordingly, we refer to a decision epoch as a time step, whose time duration depends on both the users' data requests and the ECN's status update decision, as specified in the sequel.

At a generic time step $t$, each user $n$ may request at most one data packet from the ECN, and the corresponding request details, also called query profile, can be expressed as $\mathbf R_n(t) = \left( r_n^1(t), r_n^2(t), \ldots, r_n^K(t)\right)$, with $r_n^k(t) \in \{0,1\}, \forall n \in \mathcal N, \forall k \in \mathcal K$, and $\sum\nolimits_{k = 1}^K r_n^k(t) \in \{0, 1\}$. Wherein, $r_n^k(t)=1$ if user $n$ requests the data generated by sensor $k$, and $r_n^k(t)=0$ otherwise. As such, $r_n(t)=\sum\nolimits_{k = 1}^K r_n^k(t) = 0$ means that no update packet is requested by the user in time step $t$. In this paper, we assume that the ECN has full access to the data query profiles of all users $\mathbf R(t) = \left(\mathbf R_1(t),\mathbf R_2(t), \ldots, \mathbf R_N(t)\right)$ at the beginning of each time step $t$,\footnote{If the ECN cannot completely observe these profiles, partially observable MDP (POMDP) might be used to model the status update problem. However, this is left for future work.} while it bears no prior knowledge about the request arrival rates of users or the popularity of cached items.

After obtaining the data query profiles $\mathbf R(t)$ in time step $t$, the ECN first has to decide whether or not to ask some of the sensors to update their current status, and, if any, complete the status update during SUP. Then, the ECN would respond to the users' demands, if any, with the cached packets during DDP. Similar to previous studies \cite{Caching_Multicasting_BZhou_2016,DRL_Updating_2020,IoT_Caching_2020}, we consider the ECN is able to simultaneously transmit the required data packets to the corresponding users, e.g., via multicasting, which is completed at the end of DDP.\footnote{The model can also be extended to the case where the user is allowed to request update packets generated by multiple sensors in each time step. In this case, the user devices should be upgraded so as to successfully receive data packets from multiple frequency bands simultaneously.} We denote the ECN's status update decision at time step $t$ by $\mathbf {A}(t) = \left({a}_1(t), {a}_2(t), \ldots, {a}_K(t)\right)$, where ${a}_k(t) \in \{0, 1\}, \forall k \in \mathcal K$. Particularly, ${a}_k(t) = 1$ if sensor $k$ is selected to sense the underlying environment and update its current status during the SUP, and ${a}_k(t) = 0$ otherwise. It is noteworthy that, in one time step, if there is no sensor being selected to update the status, then no SUP is included in this time step. Besides, a status update may also be generated even if there is no data requests from the users. In this case, the duration of the time step is one time slot if there is no sensor being activated to update status, and $D_u$ time slots otherwise. In this respect, as shown in Fig. \ref{Fig:System_Model_Time_Steps} (b), the duration of the time step $t$, $D(t)$, is \textit{non-uniform}, which is determined by the users' data requests and the status update decision made by the ECN, i.e.,
\begin{align} \label{Eq:Time_Duration}
D(t) \!= \!\begin{cases}
D_u \!+ \!D_d, \! \!\! \!&\textrm{if} \ \! \sum\nolimits_{n = 1}^N \! r_n(t) \! \neq 0 \&  \sum\nolimits_{k = 1}^K \! a_k(t)\! \neq \! 0\\
D_u, &\textrm{if} \ \! \sum\nolimits_{n = 1}^N \! r_n(t)\! = \!0 \& \sum\nolimits_{k = 1}^K \! a_k(t)\! \neq \! 0\\
D_d, &\textrm{if} \ \! \sum\nolimits_{n = 1}^N \! r_n(t)\! \neq  0 \& \!\sum\nolimits_{k = 1}^K \! a_k(t) \!= \! 0\\
1, & \textrm{otherwise}
\end{cases}
\end{align}
where $\sum\nolimits_{n = 1}^N r_n(t) = 0$ denotes that there is no user requesting any data packets in time step $t$, and $\sum\nolimits_{n = 1}^N r_n(t) = 1$ otherwise.

In each time step, we consider that all data packets requested by the users can be successfully transmitted from the ECN\footnote{The proposed framework and designed DRL algorithm can also be extended to cope with the error-prone channel model for the multicast transmission from the ECN to users, since in that case the transmission failure probabilities can be regarded as part of the unknown environmental dynamics.}, while at most $M \leq K$ sensors can update their status packets simultaneously without collisions over the orthogonal channels, i.e., $\sum\nolimits_{k = 1}^K a_k(t) \leq M$. Owing to the limited communication capability of sensors, we consider that the wireless channel from each sensor to the ECN is error-prone, which is presented in detail in the following subsection.

\subsection{Status update model}
During the SUP in time step $t$, we denote $\mathcal K_u(t) = \left\{ {k\left| {{a}_k(t) = 1,k \in \mathcal K} \right.} \right\}$ as the set of sensors activated by the ECN for status update, where $\sum\nolimits_{k \in \mathcal K_u(t)} a_k(t) \leq M$. In this work, the energy consumption for completing one status update is allowed to be different for different sensors. Particularly, for a generic sensor $k$, the energy consumption for environmental sensing and packet transmission are assumed to be constant and denoted by $E_{k, s}$ and $E_{k, u}$, respectively. As such, to complete one status update, the overall energy consumption of sensor $k$ is $E_k = E_{k,s} + E_{k,u}$. It is noteworthy that, since the ECN is generally powered by the power grid, its energy consumption is not a major issue for caching enabled IoT networks.

Because the wireless medium is by nature unreliable, transmissions over the spectrum may fail. We thus assume that the transmission failures of a generic sensor $k$ occur independently and identically over different transmissions with probability $p_{O}^k$, which is unknown to the ECN or sensors in advance.\footnote{We note that the channel model can also be extended to the cases where (a) the channel conditions vary depending on the number of transmitting sensors and (b) the wireless channel is modeled as a finite-state Markov chain. For these two cases, the finite-state MDP can also be formulated and hence, our established methodology for algorithm design does also apply.}
We use $z_k(t) \in \{0, 1\}$ to denote whether there is a status update successfully delivered to the ECN from sensor $k$ in time step $t$, where $z_k(t) = 1$ if the update is successfully delivered and $z_k(t) = 0$ otherwise. As such, for an activated sensor $k$ we have $\mathsf {Pr}(z_k(t) = 1 ) = 1-p_O^k$ and $\mathsf {Pr}(z_k(t) = 0 ) = p_O^k$. We note that each sensor will generate and transmit the update packet only if it is asked by the ECN. As such, we further have $z_k(t) \leq a_k(t)$ with any time step $t$.

\subsection{AoI dynamics at the ECN and users}

The metric that we adopt to evaluate the information freshness in this paper is the AoI, which is defined from the perspective of the data receiver. Particularly, we calculate the AoI at the ECN and users just before the decision moment, i.e., \textit{at the beginning of each time step}, since it is essential information for the ECN to make a well-informed decision. We assume that the maximum allowable AoI at the ECN and users is $\Delta_{\max}$, which is finite but can be arbitrarily large. The reason comes from the fact that, for real-time applications, the status update with an infinite age would be too stale to be useful. Before formally depicting the evolution of the AoI values with different sensors at each user, we first present the dynamics of the AoI for individual cached data packets from the ECN's perspective.
Specifically, for time step $t$, by denoting the AoI of sensor $k$ at the ECN with $\Delta _0^k(t)$, we can express its evolution as
{\begin{align} \label{Eq:AoI_Cache_Evolution}
&\Delta_0^k(t)= \\ \nonumber
&\begin{cases}
D(t-1), &\textrm{if} \ z_k(t-1)=1\\
\min \{\Delta_0^k(t-1)+ D(t-1), \Delta_{\max}\}, & \textrm{otherwise}
\end{cases}
\end{align}
where $z_k(t-1)=1$ indicates that during the previous time step $t-1$, an update packet generated by sensor $k$ was successfully delivered to the ECN, and $D(t-1)$ denotes the duration of the previous time step $t-1$, as specified in (\ref{Eq:Time_Duration}).

\begin{figure}[!t]
\centering \includegraphics[width=3.0in]{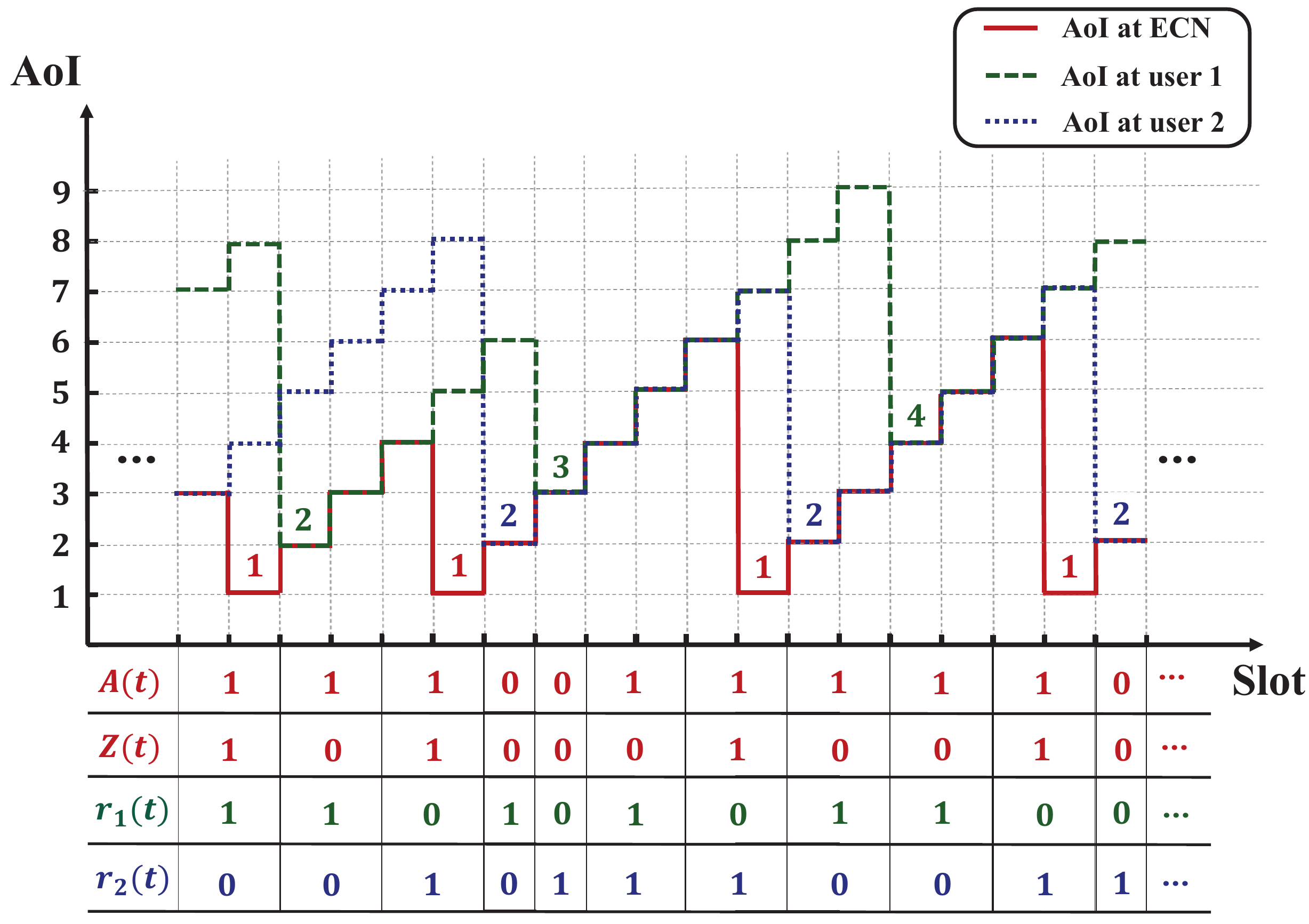}
\centering \caption{An illustration of the AoI evolution in a caching enabled IoT network, with $N=2$, $K=1$, and $D_u = D_d =1$. Wherein, the AoI evolution at the ECN, user 1 and user 2 are illustrated with the red, green and blue lines, respectively.} \label{Fig:Age_variation_Example}
\end{figure}

\begin{figure*} [!t]
\begin{align} \label{Eq:AoI_Evolution_Per_User_Per_Data}
\Delta_n^k(t) &=
\begin{cases}
D_u+D_d, &\textrm{if} \ r_n^k(t-1) z_k(t-1)=1\\
\min \{\Delta_0^k(t-1) + D_d, \Delta_{\max}\},  & \textrm{if} \ r_n^k(t-1)=1 \& \sum\nolimits_{k = 1}^K a_k(t-1)  = 0\\
\min \{\Delta_0^k(t-1) +D_u+ D_d, \Delta_{\max}\},  & \textrm{if} \ r_n^k(t-1)=1 \& \sum\nolimits_{k = 1}^K a_k(t-1) \neq 0 \& z_k(t-1)=0 \\
\min \{\Delta_n^k(t-1) + D_d, \Delta_{\max}\},  & \textrm{if} \ r_n^k(t-1)=0 \& \sum\nolimits_{n = 1}^N r_n(t-1) \neq 0 \& \sum\nolimits_{k = 1}^K a_k(t-1)  = 0 \\
\min \{\Delta_n^k(t-1) + D_u, \Delta_{\max}\},  & \textrm{if} \ \sum\nolimits_{n = 1}^N r_n(t-1) = 0 \& \sum\nolimits_{k = 1}^K a_k(t-1)  \neq 0 \\
\min \{\Delta_n^k(t-1) + 1, \Delta_{\max}\},  & \textrm{if} \ \sum\nolimits_{n = 1}^N r_n(t-1)=0 \& \sum\nolimits_{k = 1}^K a_k(t-1)  = 0 \\
\min \{\Delta_n^k(t-1)+D_u+D_d, \Delta_{\max}\}, & \textrm{otherwise}
\end{cases}
\end{align}
\hrulefill
\end{figure*}

Similarly, let $\Delta_n^k(t)$ denote the AoI of sensor $k$ perceived by user $n$. Then, its dynamics can be expressed as (\ref{Eq:AoI_Evolution_Per_User_Per_Data}), given at the top of the next page, where $r_n^k(t-1) z_k(t-1)=1$ indicates that in the previous time step $t-1$ the update data packet required by user $n$ was successfully delivered to the ECN from sensor $k$. From  (\ref{Eq:AoI_Evolution_Per_User_Per_Data}), it can be seen that the AoI at one user is jointly affected by the request behavior of all users, ECN's update decision, and update result of the corresponding data packet. To more concretely express the relationship between the AoI dynamics at the ECN and at each user, we rewrite  (\ref{Eq:AoI_Evolution_Per_User_Per_Data}) as the following equation (\ref{Eq:AoI_Evolution_Per_User_Brief}) by integrating  (\ref{Eq:Time_Duration}) and (\ref{Eq:AoI_Cache_Evolution}):
\begin{align} \label{Eq:AoI_Evolution_Per_User_Brief}
&\Delta_n^k(t) = \\ \nonumber
&\begin{cases}
\Delta_0^k(t),  & \textrm{if} \ r_n^k(t-1)=1\\
\min \{\Delta_n^k(t-1)+D(t-1), \Delta_{\max}\}, & \textrm{otherwise}
\end{cases}.
\end{align}
Intuitively, (\ref{Eq:AoI_Evolution_Per_User_Brief}) means that if in the previous time step $t-1$ a user $n$ requested a cached packet generated by sensor $k$, then the AoI of sensor $k$ at the user would be synchronous with that at the ECN at the beginning of time step $t$. Otherwise, the AoI at the user would increase at most $D(t-1)$ time slots i.e., the duration of time step $t-1$. From (\ref{Eq:AoI_Evolution_Per_User_Brief}), we note that even obtaining data from \textit{the same ECN, different AoI values} may be experienced by different users, which is unrevealed if we only focus on the dynamics of the AoI at the ECN. To better illustrate this concept, an example of the AoI evolution process is provided in Fig. \ref{Fig:Age_variation_Example}. Without loss of generality, we initialize $\Delta_0^k(1)=\Delta_n^k(1)= 0, \forall n \in \mathcal N, \forall k \in \mathcal K$.

\subsection{Problem formulation}
In the sequel, we aim at designing a dynamic status update strategy to simultaneously decrease the AoI experienced by users and reduce the energy consumed by sensors. The effect of each update decision in time step $t$ on the energy consumption of sensors is easy to be evaluated, which can be expressed as $C_E(t) = \sum\nolimits_{k = 1}^K a_k(t) E_k$. However, due to the non-uniform duration of the time steps, such an effect on the average AoI experienced by users cannot be depicted as in available studies, where the time duration between any two successive decisions is considered to be constant.

To deal with this issue, we further investigate the AoI dynamics for each user within one time step, and then amortize the AoI related cost over the included time slots. Particularly, let $i^t$ denote the $i$-th time slot in time step $t$, i.e., $i^t \in \{1, 2, \ldots, D(t)\}$, and $\Delta _{n}^k(t, i^t)$ the instantaneous AoI of sensor $k$ at user $n$ after that slot. Then, $\Delta _{n}^k(t, i^t)$ can be expressed as
\begin{align} \label{Eq:AoI_Evolution_Per_User_Brief_Slot}
\Delta_{n}^k(t, i^t) =
\begin{cases}
\Delta_n^k(t+1),  & \textrm{if} \ i^t = D(t) \\
\min \{ \Delta_n^k(t) + i^t, \Delta_{\max}\}, & \textrm{otherwise}
\end{cases}.
\end{align}
Accordingly, the average AoI related cost can be expressed as
\begin{align} \label{Eq:AoI_Cost_Per_Slot}
C_{\Delta}(t) = \sum\nolimits_{n = 1}^N {\omega_n \bar \Delta_n(t)}
\end{align}
where $\omega _n$ denotes the weight allocated to user $n$, i.e., $\omega _n \in [0, 1], \forall n \in \mathcal N$, $\sum\nolimits_{n = 1}^N \omega_n =1$, and $\bar \Delta_n(t)$ the cost associated with each user $n$, i.e.,
\begin{align} \label{Eq:AoI_Cost_Per_Slot_User}
\bar \Delta_n(t) &= \frac{1}{K} \sum\nolimits_{k = 1}^K \left({\frac{1}{D(t)}\sum\nolimits_{i^t = 1}^{D(t)}\Delta_{n}^k(t, i^t)}\right) \\ \nonumber
&= \frac{1}{K D(t)} \sum\nolimits_{k = 1}^K {\sum\nolimits_{i^t = 1}^{D(t)}\Delta_n^k(t, i^t)}
\end{align}
where $\Delta_{n}^k(t, i^t)$ is given by (\ref{Eq:AoI_Evolution_Per_User_Brief_Slot}). By amortizing the cost over time slots, we can evaluate the effects of decisions with different time durations.

To strike a balance between the average AoI experienced by users and energy consumed by sensors, we define the overall cost of the update decision made in a generic time step $t$ as
\begin{align} \label{Eq:Totoal_Cost}
C(t) = \beta_1 C_{\Delta}(t)  + \beta_2 C_{E}(t)
\end{align}
where $\beta_1$ and $ \beta_2$ are parameters used to nondimensionalize the equation and meanwhile, to make a trade-off between lowering the average AoI and reducing the energy consumption. Particularly, on one hand, given $\beta_1$, decreasing the energy consumption of sensors will gradually become the main concern as $\beta_2$ increasing. On the other hand, for the case where improving information freshness is more essential, it is preferred to adopt a larger $\beta_1$ and smaller $\beta_2$, with which the more importance is attached to reducing the average AoI experienced by users. Similar to \cite{AoI_Org_2012}, in this work we aim at designing a dynamic status update strategy to minimize the long-term average cost, i.e.,
\begin{eqnarray}   \label{Eq:Minimize_Cost}
\textbf{P}: \! \! &\mathop { \min}\limits_{ { \mathbf A^T }}& \mathop {\lim }\limits_{T \to \infty }  \frac{1}{T} \mathbb E \big[\sum\nolimits_{t = 1}^T C(t) \big] \\
&\textrm{s.t.} & \mathbf A^T = \left( {\mathbf A}(1), {\mathbf A}(2), \ldots, {\mathbf A}(T)\right) \\
&& a_k(t) \in \{0, 1\}, \forall k \in \mathcal K, \forall t  \in \{1, 2, \ldots, T\}  \label{Eq:Const_Choose} \\
&& \sum\nolimits_{k = 1}^K a_k(t) \leq M, \forall t \in \{1, 2, \ldots, T\} \label{Eq:Const_OFDMA}
\end{eqnarray}
where $\mathbf {A}^T$ denotes a sequence of update decisions made by the ECN from time step $1$ to $T$ with $\mathbf {A}(t) = \left({a}_1(t), {a}_2(t), \ldots, {a}_K(t)\right)$. Besides, constraints in (\ref{Eq:Const_Choose}) and (\ref{Eq:Const_OFDMA}) indicate that in each time step no more than $M$ sensors can be selected to update their status packets simultaneously. The ECN could make the decision in each time step by solving Problem \textbf{P}, which is a dynamic optimization problem and non-trivial to solve with standard optimization algorithms.

\section{Dueling Deep R-network based learning algorithm design}
In this section, we first formulate the dynamic status update procedure as an MDP and then, develop a dueling DRN based learning algorithm, with which the challenges brought by the unknown of transition probability and large number of state-action pairs in the formulated MDP can be addressed.

\subsection{MDP formulation}
The concerned dynamic status update procedure can be formulated as an MDP defined by a tuple $ \left(\mathbb S, \mathbb A, U\left({ \cdot , \cdot }\right)\right)$, which is depicted as follows:

\begin{itemize} \setcounter{enumi}{0}
\item \textbf{State space $\mathbb S$}: At time step $t$, the state $\mathcal S(t)$ is defined to be the combination of the AoI values at the ECN and users at the beginning of that step, i.e., $\mathcal S(t) = \left(\mathcal S_0(t), \mathcal S _1(t), \mathcal S _2(t), \ldots, \mathcal S _N(t)\right)$, where $\mathcal S _m(t) = (\Delta _m^1(t), \Delta _m^2(t), \ldots, \Delta _m^K(t)), \forall m \in \{0\} \cup\mathcal N$. Since the maximum allowable AoI at the ECN and users is finite, the state space $\mathbb S$ is finite and can be expressed as $\mathbb{S} = {{\mathbb S_0}} \times {{\mathbb S_1}} \times {{\mathbb S_2}} \times \cdots \times {{\mathbb S_{N}}}$, where ${\mathbb S_m}$, ${\forall m \in  \{0\} \cup\mathcal N}$, denotes the subspace associated with node $m$.
\item \textbf{Action space $\mathbb A$}: At the beginning of time step $t$, the ECN can activate no more than $M$ sensors to sense the underlying environment and transmit their update packets. The action space can thus be expressed as
    \begin{align} \label{Eq:Action_space}
    \mathbb{A} = \left\{ {\mathbf A\left| {\sum\nolimits_{k = 1}^K {{a_k} \le M} , {a_k} \in \left\{ {0,1} \right\}, \forall k \in \mathcal K} \right.} \right\}
    \end{align}
    where each element $\mathbf A = (a_1, a_2, \ldots, a_K)$ denotes one available status update decision.
\item \textbf{Utility function $U\left({ \cdot , \cdot }\right)$}: $U\left({ \cdot , \cdot }\right)$ is a mapping from a state-action pair $\mathcal S \times \mathbf A$  to a real number, which is used to quantify the obtained reward by choosing action $\mathbf A \in \mathbb A$ at state $\mathcal S \in \mathbb S$. In this work, the non-uniform duration of the time step is dealt with by amortizing the incurred cost in one time step over the incorporated time slots (see Section II-D). As such, for the state-action pair $\mathcal S(t)\times \mathbf A(t)$ at time step $t$, we define the utility function as $U\left({ \mathcal S(t), \mathbf A(t) }\right) = -C(t)$, where $C(t)$ is given in  (\ref{Eq:Totoal_Cost}).
\item \textbf{Transition probability $\mathsf {Pr}\left( { \cdot \left| { \cdot , \cdot } \right.} \right)$}: The transition probability defines the dynamics of the MDP, which is a mapping: $\mathbb S \times \mathbb A \times \mathbb S \to [0, 1]$, with $\sum\nolimits_{\mathcal S' \in \mathbb S} {\mathsf {Pr}\left( { \mathcal S' \left| { \mathcal S , \mathbf A } \right.} \right)}= 1, \forall \mathcal S \in \mathbb S, \forall \mathbf A \in \mathbb A$. That is, if the agent at state $\mathcal S$ performs action $\mathbf A$, then the probability of the state transition from $\mathcal S$ to $\mathcal S'$ is $\mathsf {Pr}\left( { \mathcal S' \left| { \mathcal S, \mathbf A} \right.} \right)$. Given a state-action pair $(\mathcal S, \mathbf A)$, the transition probability $\mathsf {Pr}\left( { \mathcal S' \left| { \mathcal S, \mathbf A} \right.} \right)$ is determined by the environmental dynamics, i.e., users' request arrival rates, the popularity of cached items, and the transmission failure probabilities of sensors, which is unknown to the ECN in advance.
\end{itemize}

In this work, we aim at deriving a deterministic stationary policy $\pi ^*$ that maximizes the long-term average reward with the initial state $\mathcal S(1)$, i.e.,
\begin{align} \label{Eq:Optimal_Policy}
\nonumber \pi^* &= \mathop {\arg }\limits_ \pi \max \mathop {\lim }\limits_{T \to \infty }  \frac{1}{T} \mathbb E \big[ \sum\nolimits_{t = 1}^T U\left({ \mathcal S(t), \mathbf A(t)}\right) \left| \mathcal S(1) \right. \big] \\
& \mathop  = \limits^{(a)}  \mathop {\arg }\limits_ \pi \min \mathop {\lim }\limits_{T \to \infty } \frac{1}{T} \mathbb E \big[ \sum\nolimits_{t = 1}^T C(t) \big]
\end{align}
where (a) holds when all the elements in $\mathcal S(1)$ are set to $0$. For the ECN, the deterministic and stationary status update policy is defined as follows.
\begin{definition}
A feasible deterministic and stationary policy $\pi: \mathbb S \to \mathbb A$ is defined as a mapping from the observed state $\mathcal S \in \mathbb S$ to a feasible status update action $\mathbf A \in \mathbb A$, which is irrelevant to the time step.
\end{definition}

As commonly done in available literatures (e.g., \cite{RL_1993,BZhou_Samp_Up_2019}), here we restrict our attention to stationary unichain policies, for which our formulated finite MDP is unichain, and hence, the optimal stationary policy exists, for which the achieved average reward is independent of the initial state. Comparing (\ref{Eq:Minimize_Cost}) with (\ref{Eq:Optimal_Policy}), we note that $\pi ^*$ can also be used to derive a solution to the original Problem \textbf{P}.

\begin{figure*} [!t]
\begin{align} \label{Eq:Act_Value_Fun}
R_{\pi}(\mathcal S,  \mathbf A) = \mathbb E_{\pi} \big[ \sum\nolimits_{l = 0}^\infty \left(U\left( {{\cal S}(t + l),{ \mathbf A}(t + l)}\right)- \bar U_{\pi}\right) \left| {{\cal S}(t) = \mathcal S, \mathbf A(t) =  \mathbf A} \right.\big]
\end{align}
\hrulefill
\end{figure*}

\begin{figure*} [!t]
\begin{align} \label{Eq:Act_Value_Bellman_Opt_Fun}
R_{\pi^*}(\mathcal S, \mathbf A) &= \mathbb E_{\pi *} \big[ U\left( {\mathcal S(t),\mathbf A(t)} \right) - \bar U_{\pi^*}+ \mathop {\max }\limits_{ \mathbf A'  \in \mathbb A} {R_{{\pi ^*}}}({\mathcal S}(t + 1), \mathbf A') \left| {{\mathcal S}(t) = \mathcal S, \mathbf A(t) = \mathbf A} \right. \big] \\ \nonumber
& = \sum\nolimits_{U \in \mathcal U, \mathcal S' \in \mathbb S}{\mathsf {Pr}\left( {U,{\mathcal S}'\left| {{\mathcal S},{ \mathbf A}} \right.} \right)} \left( U  - \bar U_{\pi^*} +    \mathop {\max }\limits_{\mathbf A' \in \mathbb A} {R_{{\pi ^*}}}(\mathcal S', \mathbf A') \right)
\end{align}
\hrulefill
\end{figure*}

As shown in (\ref{Eq:Optimal_Policy}), given a policy $\pi$, in each time step $t$ the action $\mathbf A(t)$ affects the obtained instantaneous reward and the state transition, which in return impacts the decisions made in subsequent time steps and the finally obtained long-term average reward. Therefore, to find the optimal strategy $\pi ^*$, it is essential to accurately estimate the long-term effect of the decision made in each time step, which is non-trivial because of the causality. In this work, we have devised a dueling DRN based status update algorithm to solve the optimization problem by combining R-learning and dueling DQN. The detailed procedure is elaborated on in the following subsection.

\subsection{DDR-DSU algorithm design}

To maximize the long-term average reward, we define the action-value function as in (\ref{Eq:Act_Value_Fun}), given at the top of the next page, in which $( \mathcal S, \mathbf A)$ denotes the initial state-action pair, $\pi$ the adopted deterministic stationary policy, and $\bar U_{\pi}$ the corresponding long-term average reward. As presented in \cite{RL_1993}, increasing the action-value evaluated with (\ref{Eq:Act_Value_Fun}) is crucial to maximize the long-term average reward and is the key to the policy improvement in RL. That is, in any state $\mathcal S$, changing $\pi$ to choose an action $\mathbf A$ that results in an improvement of $R_{\pi}\left(\mathcal S, \mathbf A\right)$ would also improve the average $\bar U_{\pi}$. For this action-value function $R_{\pi}\left(\mathcal S, \mathbf A\right)$, the Bellman optimality equation can be expressed as (\ref{Eq:Act_Value_Bellman_Opt_Fun}), given at the top of this page, in which ${\mathsf {Pr}\left( {{U, \mathcal S}'\left| {{\mathcal S},{\mathbf A}} \right.} \right)}$ denotes the probability that the reward-state pair $(U, \mathcal S')$ is observed by the agent, if it performs action $\mathbf A$ at state $\mathcal S$. Here, $\mathcal U$ denotes the reward set with a finite number of elements \cite{RL_Introduction}. If the probability is known, the Bellman optimality equations are in essence a system of equations with $\left| {\mathbb S \times \mathbb A} \right|$ unknowns, where $\left| {\cdot} \right|$ represents the cardinality of a set. And, for a finite MDP, we can find its solution by utilizing model-based RL algorithms, e.g., dynamic programming \cite{RL_Introduction}.

In this work, to solve the equations in (\ref{Eq:Act_Value_Bellman_Opt_Fun}), two challenges need to be addressed. On the one hand, the transition probability cannot be regarded as a prior knowledge due to the unknown pattern of users' requests as well as the unknown transmission failure probabilities of sensors. As such, a model-free learning algorithm should be adopted. On the other hand, the number of state-action pairs $\left| {\mathbb S \times \mathbb A} \right| = \left| { \mathbb A} \right|*\left(\left(\Delta_{max}\right)^{K*(N+1)}+1\right)$ increases exponentially with respect to the number of sensors and users. Therefore, standard tabular model-free algorithms, e.g., classical Sarsa as well as Q-learning introduced in \cite{RL_Introduction}, and R-learning proposed in \cite{RL_1993}, do not apply due to the curse of dimensionality.

Actually, when the number of state-action pairs is large, it is impossible to update and record the value for all of them, not only due to the memory needed for the large table (e.g., Q-table), but also the time and data needed to estimate the action-values accurately, and hence function approximation is necessary to be implemented \cite{RL_Introduction}. By combining RL with deep neural networks (i.e., DRL), the seminal work \cite{DQN_Nature_Letter} developed a novel DRL algorithm, named DQN, and then various DRL algorithms with individual improvements followed suit \cite{DQN_Variations_Ranbow}, which were widely adopted to address emerging engineering issues in modern networks \cite{DRL_Surrvey_Tao}. For the standard DQN and its variants, the objectives are similar, i.e., \textit{maximizing the discounted long-term cumulative reward}. However, for many engineering problems in practice as well as ours, the main concern is \textit{to maximize the long-term average reward instead of the discounted one}, since the interaction between the agent and environment will continue infinitely, and the rewards obtained in different time steps are equally important to the agent \cite{RL_1993}. As such, when implementing these DRL algorithms to solve our problem, there are two main limitations. First, the obtained average reward is less stable during the learning process, e.g., as presented in \cite{DQN_Nature_Letter,DDQN_2016,Priority_2016}. Second, to achieve a better performance, the discount factor (a hyperparameter) should be sophisticatedly tuned, which is non-trivial since \textit{a larger discount factor does not always bring a better performance}, e.g., see the simulation settings in recent work \cite{DF_Tuning_1,DF_Tuning_2,DF_Tuning_3}.

\begin{figure*} [!t]
\begin{align} \label{Eq:State_Value_Fun}
V_{\pi}(\mathcal S) = \mathbb E_{\pi} \big[ \sum\nolimits_{l = 0}^\infty \left(U\left( {{\cal S}(t + l),\pi({\cal S}(t + l))}\right)- \bar U_{\pi}\right)\left| {{\cal S}(t) = \mathcal S} \right. \big]
\end{align}
\hrulefill
\end{figure*}

To this end, rather than focusing on the tabular RL algorithm design as in our previous study \cite{CXu_Updating_2020}, we restrict our attention on DRL algorithms so as to circumvent the curse of dimensionality, which arises from the extremely large number of state-action pairs. Particularly, we aim at deriving a deterministic stationary policy to maximize the achieved long-term average reward, by combining the DQN and its variants with R-learning, which is a tabular model-free RL algorithm devised to maximize the undiscounted reward for continuing decision-making problems \cite{RL_1993}. Taking the dueling DQN \cite{Dueling_DQN_Org} as an example\footnote{The main motivation behind choosing dueling DQN, rather than the standard DQN, is that it can learn which states are valuable, without learning the effect of every action for each state and hence, the performance and training speed can be improved\cite{Dueling_DQN_Org}. We note that with our proposed design framework, R-learning could be also combined with other DRL algorithms. One of those is the deep R-network, a combination of the standard DQN and R-learning, whose pseudo-code is presented in Appendix \ref{Append:DR_DDU}.}, we devise the DDR-DSU algorithm by combining it with R-learning. Particularly, by introducing the dueling network architecture, we have to simultaneously estimate the state-value function $V_{\pi}(\mathcal S)$ and advantage function $G_{\pi}(\mathcal S, \mathbf A)$ \cite{Dueling_DQN_Org}, instead of directly estimating the action-value function as in the standard DQN \cite{DQN_Nature_Letter}. For our problem, $V_{\pi}(\mathcal S)$ can be defined as (\ref{Eq:State_Value_Fun}), given at the top of the next page, where the function $\pi(\cdot)$ denotes the mapping from a state $\mathcal S$ to an action $\mathbf A$ with policy $\pi$. Besides, to address the issue of identifiability and instability, the advantage function $G_{\pi}(\mathcal S, \mathbf A)$ is designed to satisfy the following equation \cite{Dueling_DQN_Org}
\begin{align} \label{Eq:Advantage_Fun}
\nonumber R_{\pi}(\mathcal S, \mathbf A) &= V_{\pi}(\mathcal S) + \\
&\left(G_{\pi}(\mathcal S, \mathbf A)- \frac{1}{\left| { \mathbb A} \right|}\sum\nolimits_{\mathbf A' \in \mathbb A} {G_{\pi}(\mathcal S, \mathbf A')}\right).
\end{align}

The architecture of our proposed DDR-DSU algorithm is illustrated in Fig. \ref{Fig:Fig2}, which consists of three key components: 1) dueling deep R-network (DDRN), 2) experience replay, and 3) target dueling deep R-network (TDDRN). For DDRN, two streams are constructed with two independent artificial neural networks (ANNs), which are parameterized by different parameters, i.e., $\mathbf {\theta_1}$ and $\mathbf {\theta_2}$, and used to approximate the state-value function and advantage function, respectively. Accordingly, for each state-action pair $\mathcal S \times \mathbf A$, the action-value function in (\ref{Eq:Advantage_Fun}) can be approximated as follows
\begin{align} \label{Eq:Act_Value_Fun_Approx}
R \left(\mathcal S,  \mathbf A; \mathbf {\theta_1}, \mathbf {\theta_2}\right) &= V\left(\mathcal S; \mathbf {\theta_1}\right) + \\ \nonumber
&\left(G\left(\mathcal S, \mathbf A; \mathbf {\theta_2}\right)- \frac{1}{\left| { \mathbb A} \right|}\sum\nolimits_{\mathbf A' \in \mathbb A} {G\left(\mathcal S, \mathbf A'; \mathbf {\theta_2}\right)}\right).
\end{align}
Here, a replay buffer is introduced to allow the agent to store the observed experiences for further processing (i.e., experience replay), which reduces the strong correlations between samples and enhances the data efficiency \cite{Introduction_DRL}. By replaying minibatches of experiences drawn from the replay buffer, temporal-difference (TD) errors are calculated to update the network parameters $\mathbf {\theta_1}$ and $\mathbf {\theta_2}$, as well as the estimated average reward $\bar U$, which is essential to achieve the policy maximizing the long-term average reward. Besides, to stabilize the learning process, a separate target network is also introduced \cite{DQN_Variations_Ranbow}, i.e., TDDRN with parameters $\mathbf {\theta_1^{-}}$ and $\mathbf {\theta_2^{-}}$. Wherein, the parameters of TDDRN will be periodically updated with those of DDRN, i.e., set $\mathbf {\theta_1^{-}} = \mathbf {\theta_1}$ and $\mathbf {\theta_2^{-}} = \mathbf {\theta_2}$.

\begin{figure} [!t]
\centering
\leavevmode \epsfxsize=3.0in  \epsfbox{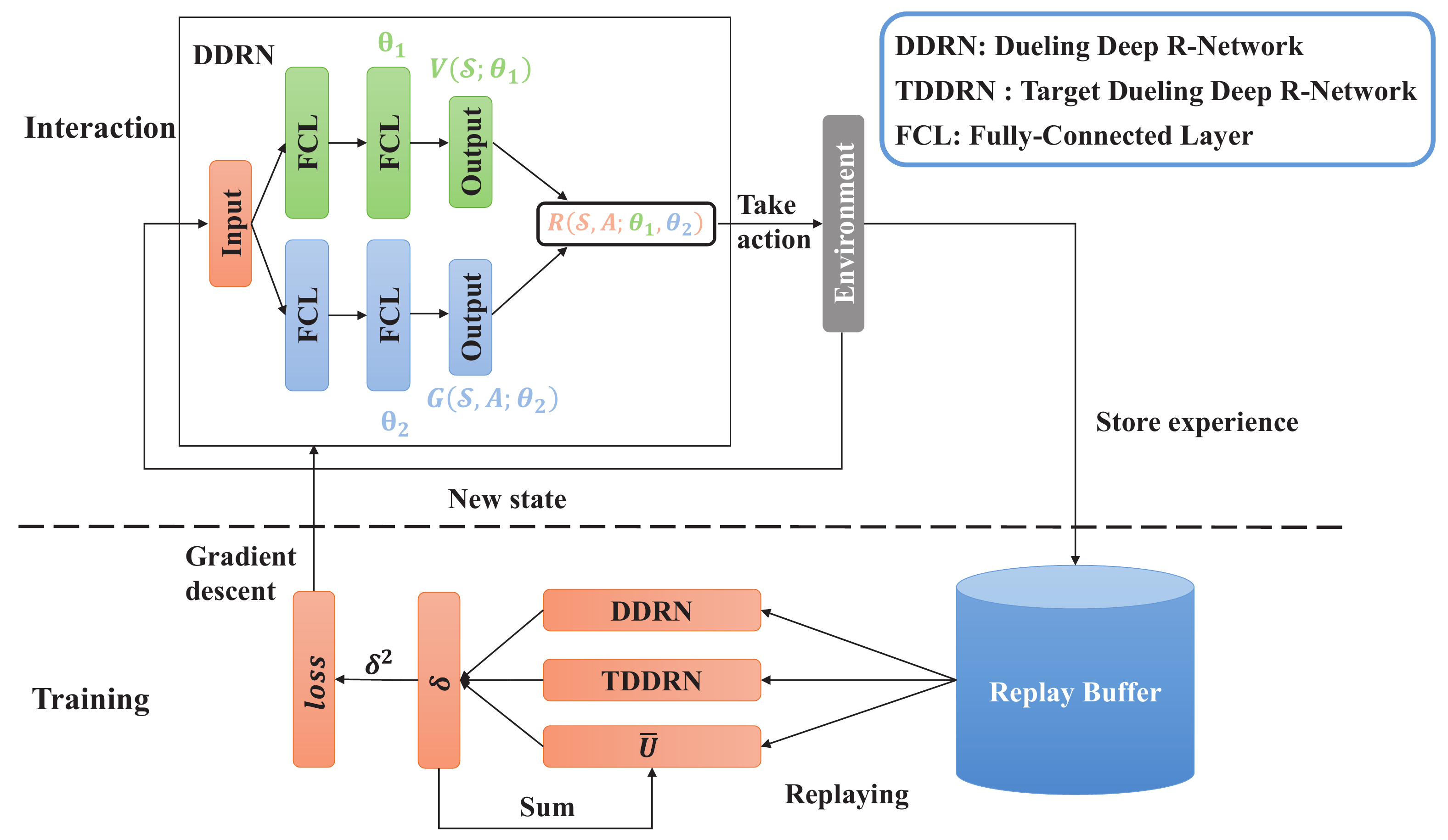}
\centering \caption{Architecture of our proposed DDR-DSU algorithm.} \label{Fig:Fig2}
\end{figure}

\begin{algorithm}[!t]
\caption{Dueling deep R-network based dynamic status update (DDR-DSU) algorithm.}
\begin{algorithmic}[1] \label{Alg:DDR_DDU_algorithm}
\STATE \textbf{Initialization:}
\STATE Initialize the experience replay buffer $\mathbb D$, average reward $\bar U$, DDRN parameters $\mathbf {\theta_1}$ and $\mathbf {\theta_2}$, TDDRN parameters $\mathbf {\theta_1^{-}} = \mathbf {\theta_1}$ and $\mathbf {\theta_2^{-}} = \mathbf {\theta_2}$, maximum number of training steps $T_{max}$, $t = 1$, and $\mathcal S(t) = \{0, 0, \ldots, 0\}$.
\STATE \textbf{Go into a loop:}
\FOR{$t < T_{\max} $}
\STATE \textbf{Action selection:} Choose an action $\mathbf A(t)$ according to the probability distribution in (\ref{Eq:Prob_Dis_E_Greddy}). \label{Alg:Repeat_Begin}
\STATE \textbf{Acting and observing:} Take action $\mathbf A(t)$, obtain a reward $U\left({ \mathcal S(t), \mathbf A(t)}\right)$, and observe a new state $ \mathcal S(t+1)$.
\STATE \textbf{Refreshing replay buffer:} Store the new experience tuple $\left(\mathcal S(t), \mathbf A(t), U\left({ \mathcal S(t), \mathbf A(t)}\right), \mathcal S(t+1)\right)$ into the replay buffer $\mathbb D$. \label{Alg:Repeat_End}
\IF {$t > T_s$}
\STATE \textbf{Replaying and updating (Training):}
\STATE Uniformly sample a minibatch of $D_b$ tuples from the replay buffer $\mathbb D$ and calculate the TD error for each sampled tuple with (\ref{Eq:Target_Calculate}) and (\ref{Eq:TD_Error}).
\STATE Update average reward $\bar U$ with (\ref{Eq:Ave_reward_Update}).
\STATE Perform a gradient descent step on loss $L\left( \mathbf {\theta_1}, \mathbf {\theta_2} \right)$ given in (\ref{Eq:Loss_Function}) with respect to $\mathbf {\theta_1}$ and $\mathbf {\theta_2}$ utilizing (\ref{Eq:Net_Par_Updating}), respectively.
\IF {$mod(t, T_0) = 0$}
\STATE \textbf{Updating target network:} Set the parameters $\mathbf {\theta_1^{-}} = \mathbf {\theta_1}$ and $\mathbf {\theta_2^{-}} = \mathbf {\theta_2}$.
\ENDIF
\ENDIF
\STATE Set $t = t+1$.
\ENDFOR
\STATE \textbf{Output:} Set the parameters $\mathbf {\theta_1^{-}} = \mathbf {\theta_1}$ and $\mathbf {\theta_2^{-}} = \mathbf {\theta_2}$ and output TDDRN.
\label{Alg1:Iteration_End}
\end{algorithmic}
\end{algorithm}

The details of our proposed DDR-DSU algorithm are presented in Algorithm \ref{Alg:DDR_DDU_algorithm}. At the beginning of DDR-DSU, the experience replay buffer $\mathbb D$ is cleared out, the two sets of parameters of DDRN (i.e., $\mathbf \theta_1$ and $\mathbf \theta_2$) are randomly initialized, and the parameters in TDDRN are set as $\mathbf {\theta_1^{-}} = \mathbf {\theta_1}$ and $\mathbf {\theta_2^{-}} = \mathbf {\theta_2}$. Besides, all elements of the initial state $\mathcal S(1)$ are set to $0$.
When the initialization is completed, the algorithm goes into a loop. At each iteration $t$, we first choose an action $\mathbf A(t)$ from the action space $\mathbb A$ based on the current state $\mathcal S(t)$ by resorting to the approximation of the action-value function $R \left(\mathcal S,  \mathbf A; \mathbf {\theta_1}, \mathbf {\theta_2}\right)$ with parameters $\mathbf {\theta_1}$ and $\mathbf {\theta_2}$, as in  (\ref{Eq:Act_Value_Fun_Approx}).
To balance the exploration and exploitation, we adopt the $\varepsilon$-greedy policy here, i.e., choosing an action $\mathbf A(t)$  from the space $\mathbb A$ with the following probability distribution
\begin{align}   \label{Eq:Prob_Dis_E_Greddy}
\mathsf {Pr} \left( { \mathbf A(t) \left| { \mathcal S(t)} \right.} ; \mathbf {\theta_1}, \mathbf {\theta_2} \right)
= \begin{cases}
1- \frac{\left| { \mathbb A} \right|-1}{\left| { \mathbb A} \right|}\varepsilon, & \text{{$\mathbf A(t) = \mathbf A^*$}}\\
\frac{\varepsilon}{\left| { \mathbb A} \right|}, & \text{$\mathbf A(t) \neq \mathbf A^*$}
\end{cases}
\end{align}
with $\varepsilon$ belonging to $[0, 1]$ and
\begin{align}   \label{Eq:Best_Action}
\mathbf A^* = \mathop {\arg }\limits_{\mathbf A \in \mathbb A} \max{ R \left(\mathcal S(t),  \mathbf A; \mathbf {\theta_1}, \mathbf {\theta_2}\right) }.
\end{align}
After that, we would conduct the action $\mathbf A(t)$, obtain a reward $U\left({ \mathcal S(t), \mathbf A(t)}\right)$, and observe a new state $\mathcal S(t+1)$. Then, the corresponding experience tuple $\mathcal F = \left(\mathcal S(t), \mathbf A(t), U\left({ \mathcal S(t), \mathbf A(t)}\right), \mathcal S(t+1)\right)$ is stored into the experience replay buffer $\mathbb D$ for future replay. When the buffer $\mathbb D$ is full, the oldest experience tuple is discarded.

After the iteration time $t$ is larger than a predefined threshold $T_s$, i.e., the number of experience tuples stored in $\mathbb D$ is larger than $T_s$, experience replaying and parameters updating will be conducted. Specifically, we first randomly sample a minibatch of $D_b$ experience tuples, which can be briefly expressed as $\mathbb D_b = \{\mathcal F_1, \mathcal F_2, \ldots, \mathcal F_{D_b}\}$, with $\mathcal F_i = \left(\mathcal S_i, \mathbf A_i, U_i, \mathcal S_i'\right), \forall \mathcal F_i \in \mathbb D_b$. Then, for each sampled experience tuple $\mathcal F_i$, the corresponding target and TD error can be calculated as
\begin{align} \label{Eq:Target_Calculate}
\hat R_i = U_i- \bar U + \mathop {\max }\limits_{\mathbf A' \in \mathbb A} {R}\left(\mathcal S_i',  \mathbf A'; \mathbf {\theta_1^-}, \mathbf {\theta_2^-}\right)
\end{align}
and
\begin{align}   \label{Eq:TD_Error}
\delta_i =  \hat R_i - {R}\left(\mathcal S_i,  \mathbf A_i; \mathbf {\theta_1}, \mathbf {\theta_2}\right)
\end{align}
respectively, where ${R}\left(\cdot,  \cdot  \ ; \mathbf {\theta_1}, \mathbf {\theta_2}\right)$ and ${R}\left(\cdot,  \cdot \ ; \mathbf {\theta_1^-}, \mathbf {\theta_2}^-\right)$ denote the action-value function approximated by respectively utilizing the DDRN and TDDRN.

With the TD errors for these sampled experiences in one minibatch, the estimation of the corresponding long-term average reward $\bar U$ and the parameters $\mathbf {\theta_1}$  as well as $\mathbf {\theta_2}$ can be updated. Particularly, we update $\bar U$ with the sum of TD errors of the sampled experiences in one minibatch
\begin{align}   \label{Eq:Ave_reward_Update}
\bar U = \bar U + \alpha_0 \sum\nolimits_{i = 1}^{D_b} {\delta_i}
\end{align}
where $\alpha_0$ is the learning rate. Besides, the parameters can be updated as
\begin{align}   \label{Eq:Net_Par_Updating}
\mathbf \theta_l = \mathbf \theta_l - \alpha_l \nabla_{\theta_l} L\left( \mathbf {\theta_1}, \mathbf {\theta_2} \right), \forall l \in \{1,2\}
\end{align}
where $ \alpha_l$ represents the learning rate for updating parameters $\mathbf \theta{_l}$, $\forall l \in \{1,2\}$, and $L\left( \mathbf {\theta_1}, \mathbf {\theta_2} \right)$ the loss function at each iteration, i.e.,
\begin{align}   \label{Eq:Loss_Function}
L\left( \mathbf {\theta_1}, \mathbf {\theta_2} \right) = \frac{1}{D_b} \sum\nolimits_{i =1}^{D_b} (\delta_i) ^2.
\end{align}
During the replaying and updating phase, the parameters of the target network $\mathbf {\theta_1^{-}}$ and $\mathbf {\theta_2^{-}}$ are respectively replaced by those of DDRN (i.e., $\mathbf {\theta_1}$ and $\mathbf {\theta_2}$) every $T_0$ iterations.

\begin{remark}
Time complexity. The time complexity of training an ANN is determined by the number of operations in each iteration during the update. For a fully-connected ANN just with $q_{I}$ input and $q_{O}$ output neurons, the time complexity is $\mathcal O(q_{I}q_{O})$ \cite{Wu_2016_CVPR}. Furthermore, if the fully-connected ANN consists of one input, one output as well as $H$ hidden layers, whose numbers of neurons are respectively denoted by $q_{I}$,  $q_{O}$, and $q_{h}$, $\forall h \in \{1, 2, \ldots, H\}$, then its corresponding time complexity in each iteration can be expressed as $\mathcal O(q_{I}q_{1}+\sum\nolimits_{h =1}^{H-1}q_{h}q_{h+1}+q_{H}q_{O})$\cite{Complexity_Ana_UAV}. As such, for our developed DDR-DSU algorithm with two separate ANNs, the time complexity of each update can be expressed as
\[\mathcal O(( \mathbb S)_D(q_{1}^1+q_{1}^2)+\sum\nolimits_{h =1}^{H-1}(q_{h}^1q_{h+1}^1+q_{h}^2q_{h+1}^2)+q_{H}^1*1+q_{H}^2*\left| { \mathbb A} \right|)\]
where $( \mathbb S)_D = K*(N+1)$ represents the dimension of the state space $\mathbb S$, i.e., the number of input neurons, and $q_h^1$ and $q_h^2$, $\forall h \in \{1, 2, \ldots, H\}$, respectively denote the numbers of neurons in hidden layer $h$ of the two ANNs, utilized to approximate the state-value and advantage functions.
\end{remark}

When the algorithm is terminated, we obtain an approximated action-value function mapping from each state-action pair to a real number by utilizing TDDRN, i.e., $R \left(\mathcal S, \mathbf A; \mathbf {\theta_1^{-}}, \mathbf {\theta_2^{-}}\right)$, $\forall \mathcal S \in \mathbb S, \mathbf A \in \mathbb A$. As scuh, we can obtain an approximate solution to Problem \textbf{P} by accessing the trained TDDRN and choosing the action bringing the maximum action-value at the beginning of each time step. To avoid the heavy computation iteration and cold-start problem, when implementing our proposed algorithm in practice, it is recommended to conduct offline training and online inference. Actually, as demonstrated in the Section IV, during our simulations, it only takes the ECN about 1.41 ms to generate an action with the trained model (i.e., complete one forward pass) for the scenario with 8 sensors and 48 users, allowing the system to insert real-time decisions on status update in practice. It shall be noted that the convergence analysis of our proposed algorithm is left untouched because this is still an open problem for DRL algorithms. Nonetheless, as demonstrated by simulation results in the next section, the convergence of our proposed DRL algorithms can be steadily achieved.

\section{Simulation results} \label{Sec:Simulation}

In this section, we conduct simulations to evaluate the performance of our proposed algorithm, where TensorFlow is utilized to implement our proposed DDR-DSU and the baseline DRL algorithms. We first present the simulation setting consisting of the scenario related parameters and hyperparameters adopted by DDR-DSU and baseline policies. Then, we respectively evaluate the convergence and effectiveness, in terms of achieved long-term average reward, of the proposed algorithm in different environments.

\subsection{Simulation setting}

For the simulation scenario, we consider an IoT network consisting of $N$ users, one ECN, and $K=8$ sensors. We set both $D_u$ and $D_d$ to one time slot whose duration is 1 s. In different environments, $N$ can be set to different values. The number of orthogonal channels is set as $M=\frac{K}{2}$. To simulate the data requests of users, we consider that the requests of each user arrive at the ECN according to an independent Bernoulli process with parameter $P_n$ and with the same popularity to all sensors, i.e., in each time step $t$, we have $\mathsf {Pr}(r_n^k(t)=1) = \frac{1}{K}P_n, \forall n \in \mathcal N$, and $\mathsf {Pr}\left( \sum\nolimits_{k =1}^{K} r_n^k(t) = 0\right) = 1-P_n$. Here, we set $P_n = 0.6$, $\forall n \in \{1, 2, \ldots, N\}$. For each sensor, the adopted transmission power is 10 mW, and the energy consumption for sensing $E_s$ is the same as that for data transmission\cite{EC_Sensing}. The energy consumption related cost $C_E$ is evaluated in units of mJ. Besides, to make sensors differentiated, the transmission failure probability of a generic sensor $k$ is set as $p_O^k = 0.025*\left\lceil {\frac{k}{2}} \right\rceil $, where $\left\lceil {\cdot} \right\rceil$ denotes the operation of rounding up the number to an integer, e.g., $p_O^3 = 0.025*\left\lceil {\frac{3}{2}} \right\rceil = 0.05$. Meanwhile, the maximum AoI, $\Delta_{\max}$, is set as $\Delta_{\max} = 10K(D_u+D_d)$, and the user weight factor is $\omega_n =1/N$, $\forall n \in \{1, 2, \ldots, N\}$. The average reward $\bar U$ in Algorithm \ref{Alg:DDR_DDU_algorithm} is initialized as $-(\beta_1 \Delta_{\max}+\beta_2M(E_s+E_u))$. Unless otherwise specified, all the setting mentioned here is the default.

To evaluate the effectiveness of our proposed algorithm, we compare its performance with the following five baseline algorithms and policies:
\begin{itemize} \setcounter{enumi}{0}
\item DR-DSU: Deep R-network based dynamic status update algorithm, a combination of the standard DQN and R-learning, whose pseudo-code is given in Appendix \ref{Append:DR_DDU}.
\item DDQ-DSU: Dueling DQN based dynamic status update algorithm, where the dueling architecture is introduced into the standard DQN \cite{Dueling_DQN_Org}. Its original aim is to maximize the discounted long-term cumulative reward instead of the average one.
\item DQ-DSU: DQN based dynamic status update algorithm, where the standard DQN \cite{DQN_Nature_Letter} is utilized. Its original aim is to maximize the discounted long-term cumulative reward instead of the average one.
\item Greedy policy: With this policy, the ECN will ask sensors with the stalest $M$ cached data packets to update their current status at the beginning of each time step.
\item Random policy: With this policy, the ECN will randomly choose an available action (i.e., $\sum\nolimits_{k = 1}^K a_k(t) \leq M$) at the beginning of each time step.
\end{itemize}

To conduct the simulation, we used an NVIDIA GPU of GeForce RTX 2080 Ti. The CPU is Intel(R) Core(TM) i9-9900K@3.60GHz with 64 GB RAM. The software environment we adopted is TensorFlow 2.0 with Python 3.7 on Ubuntu 16.04 LTS. For the dueling networks (i.e., dueling deep R-network and dueling deep Q-network) as well as the corresponding target networks, we utilize two independent fully-connected ANNs, each of which is with one input, one output and two hidden layers, to approximate the state-value function and advantage function, respectively. Wherein, 128 neurons are used for both the two hidden layers, while for the output layers producing the state-value and advantage values, there are 1 and $\left| { \mathbb A} \right|$ neurons, respectively. For the DQN without the dueling architecture, we also utilized a fully-connected ANN with 4 layers. For a fair comparison, there are 256 neurons for each hidden layer and $\left| { \mathbb A} \right|$ neurons for the output layer to approximate action-values. For all algorithms, the input to ANNs, i.e. state, is normalized with $\Delta_{\max}$, and the gradients are clipped to make their norm less than or equal to 10 as in \cite{Dueling_DQN_Org}. To perform exploration and exploitation, $\varepsilon$ is annealed linearly from 1.0 to 0.01 and keeps fixed after the replay memory is full\cite{DQN_Nature_Letter}. The setting of other hyperparameters is summarized in Table \ref{Tab:Hyperparameters}.

\begin{table}[!t]
\renewcommand{\arraystretch}{1.3}
\caption{Setting of Hyperparameters.}
\label{Tab:Hyperparameters}
\centering
\begin{tabular}{ l  l  } \toprule
\textbf{Hyperparameter and description} & \textbf{{Setting}} \\
\midrule
Learning rate for parameters of ANNs, $\alpha_1, \alpha_2$                                            & $5*10^{-4}$                                                 \\
Learning rate for average reward, $\alpha_0$                                                              & $5*10^{-4}$                          \\
Minibatch size, $D_b$                                                                                         & 64                                             \\
Replay memory size, $\left| { \mathbb D} \right|$                                                         & $2*10^5$                                                \\
Replay start time, $T_s$                                                                                 & $5*10^4$                                        \\
Target network update frequency, $T_0$                                                                            & $2*10^3$                                             \\
Activation function                                                                                       & ReLU \cite{DQN_Nature_Letter}
\\
Optimizer                                                                                                 & Adam \cite{Adam_Ref}
\\
Weights initializer                                                                                       & He \cite{He_Ref}
\\
\toprule
\end{tabular}
\end{table}

\subsection{Convergence comparison}

We first investigate the convergence behavior of different DRL based dynamic status update algorithms, i.e., DDR-DSU, DR-DSU, DDQ-DSU, and DQ-DSU. For DQN based algorithms, i.e., DDQ-DSU, and DQ-DSU, to see the effect of the adopted discount factor $\gamma$ on the achieved average reward, independent simulations are conducted when $\gamma$ is set to 0.9, 0.95, and 0.99, respectively. To test the convergence of each algorithm, we periodically extract the trained target network to make decisions, i.e., executing the evaluation, during which, as in \cite{DQN_Nature_Letter}, the $\varepsilon$-greedy policy with $\varepsilon = 0.05$ is adopted. Here, the evaluation is performed every $2*10^3$ training steps, after the number of experience tuples stored in replay buffer is larger than $T_s$. The average reward is calculated by interacting with the environment for $10^4$ times, i.e., there are $10^4$ decisions being made in one evaluation. $T_{max}$ is set as $T_{max} = T_s + 120*2*10^3 = 2.9*10^5$, i.e., the learning procedure will be completed after the evaluation is performed 120 times. For each algorithm, all simulation results are obtained by averaging over 6 independent runs with different seeds, while, for fair comparison, the same seed is adopted for all algorithms in one run. The simulation results are shown in Figs. \ref{Fig:Convergence_24User} and \ref{Fig:Convergence_40User}, where the darker (solid or dotted) line shows the average value over runs and the shaded area is obtained by filling the interval between the maximum and minimum values over runs.

\begin{figure} [!t]
\centering
\subfigure[] {\leavevmode \epsfxsize=3.0in  \epsfbox{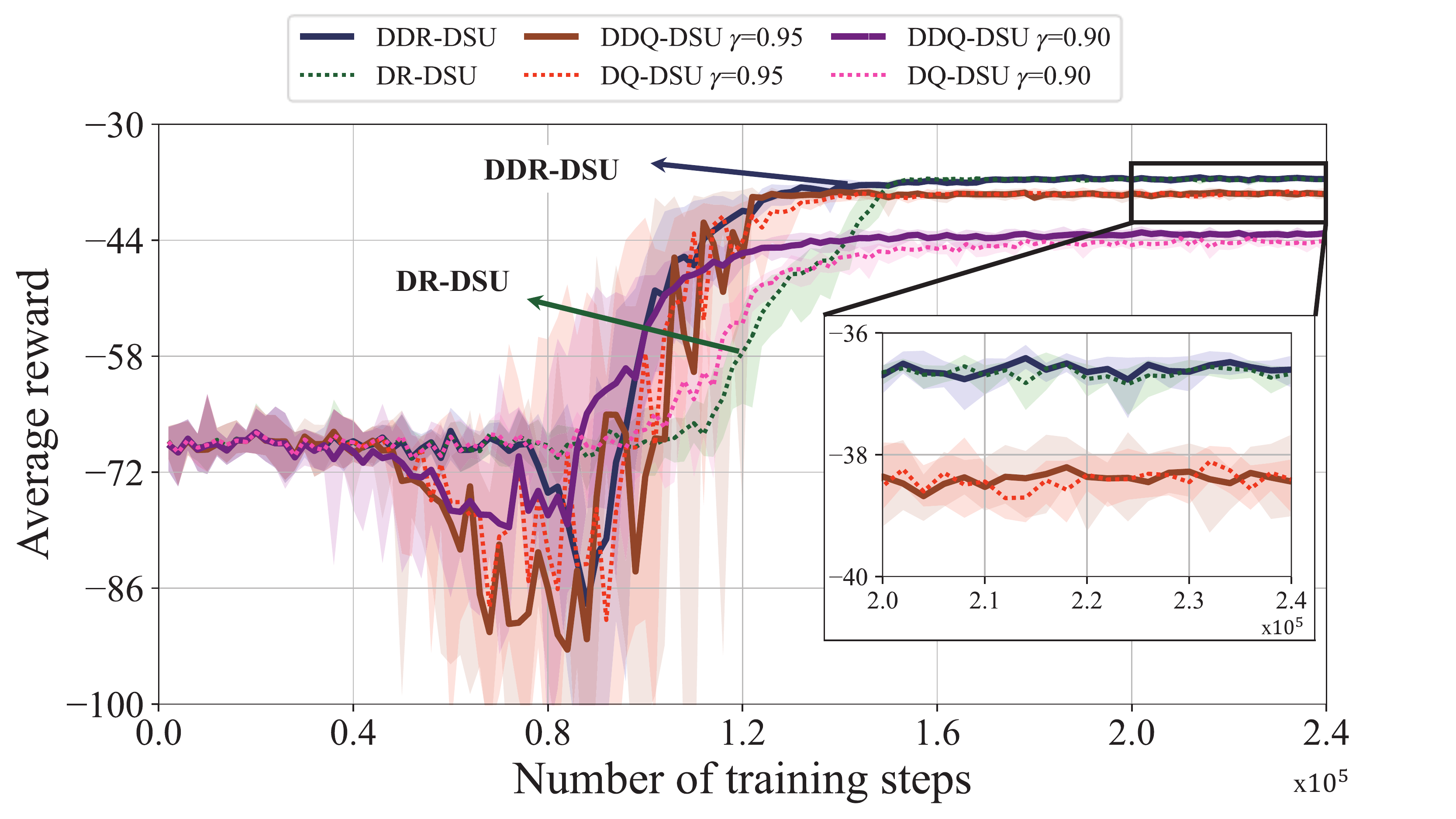}}
\subfigure[] {\leavevmode \epsfxsize=3.0in  \epsfbox{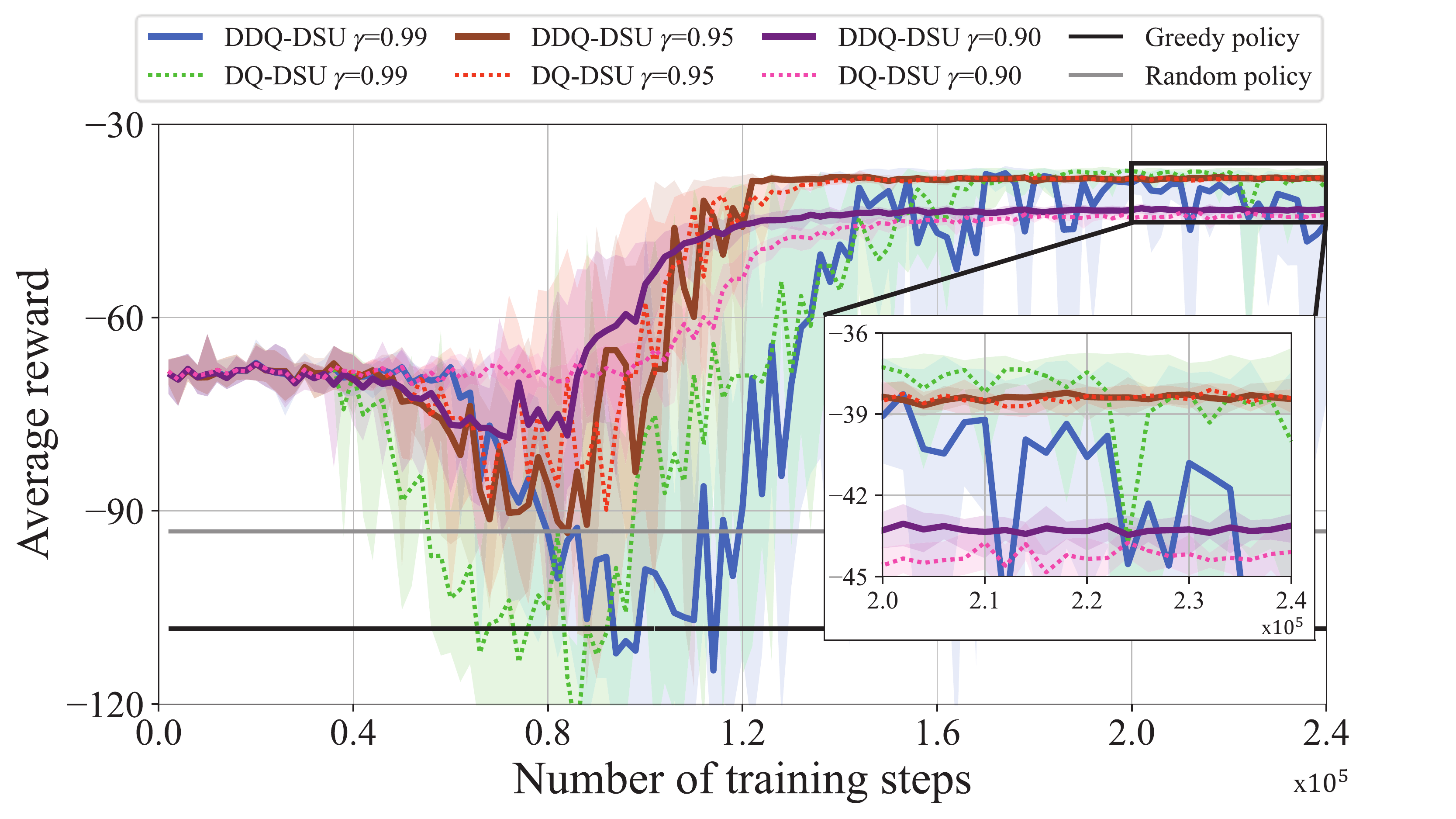}}
\centering \caption{Convergence comparison of: (a) our proposed DDR-DSU and baseline DRL algorithms; (b) baseline policies and DRL algorithms with different discount factor $\gamma$, where $N=24$ and $\beta_1 = \beta_2 =1$. Here, training refers to the operation of replaying and updating.} \label{Fig:Convergence_24User}
\end{figure}

\begin{figure} [!t]
\centering
\subfigure[] {\leavevmode \epsfxsize=3.0in  \epsfbox{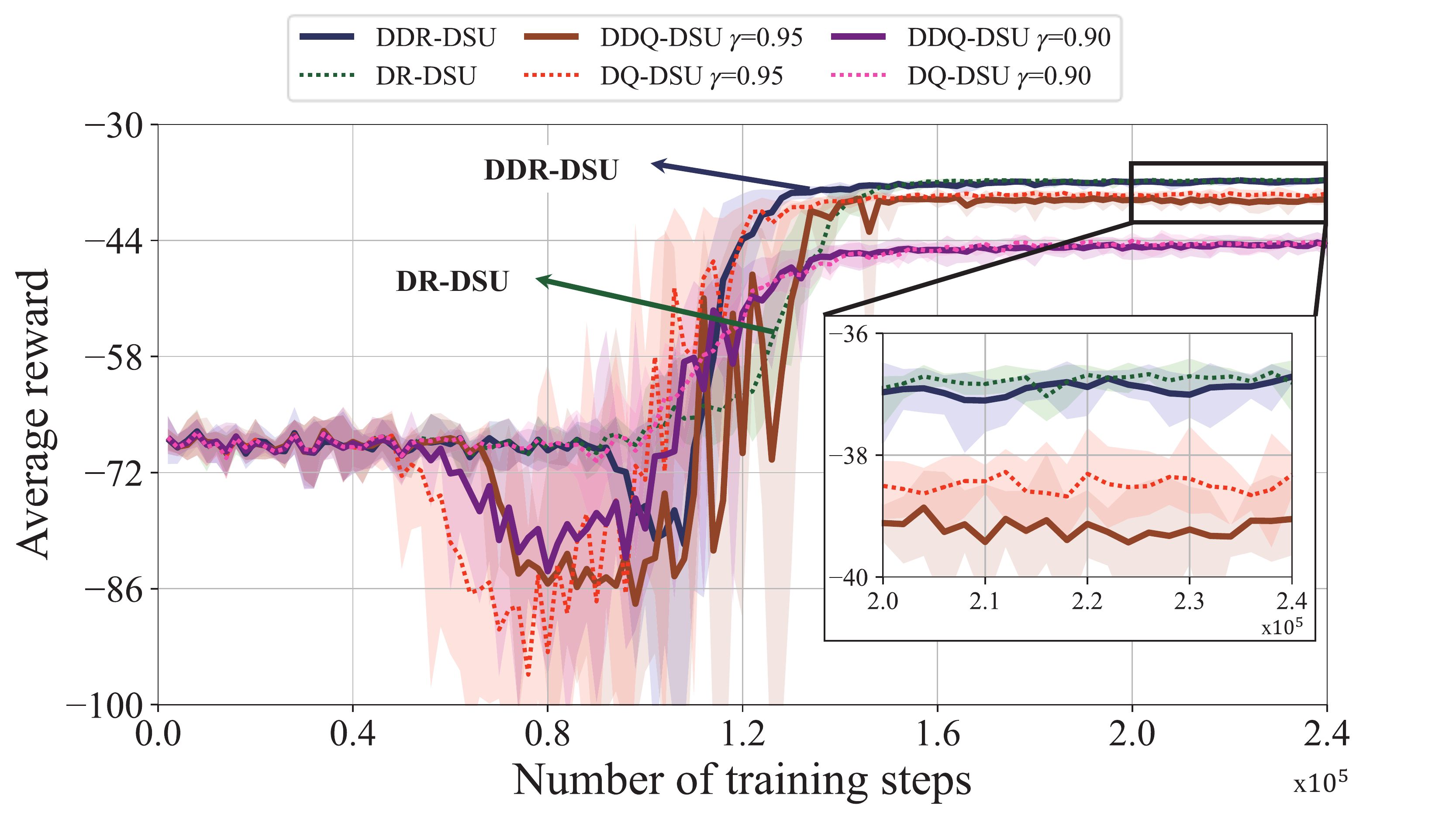}}
\subfigure[] {\leavevmode \epsfxsize=3.0in  \epsfbox{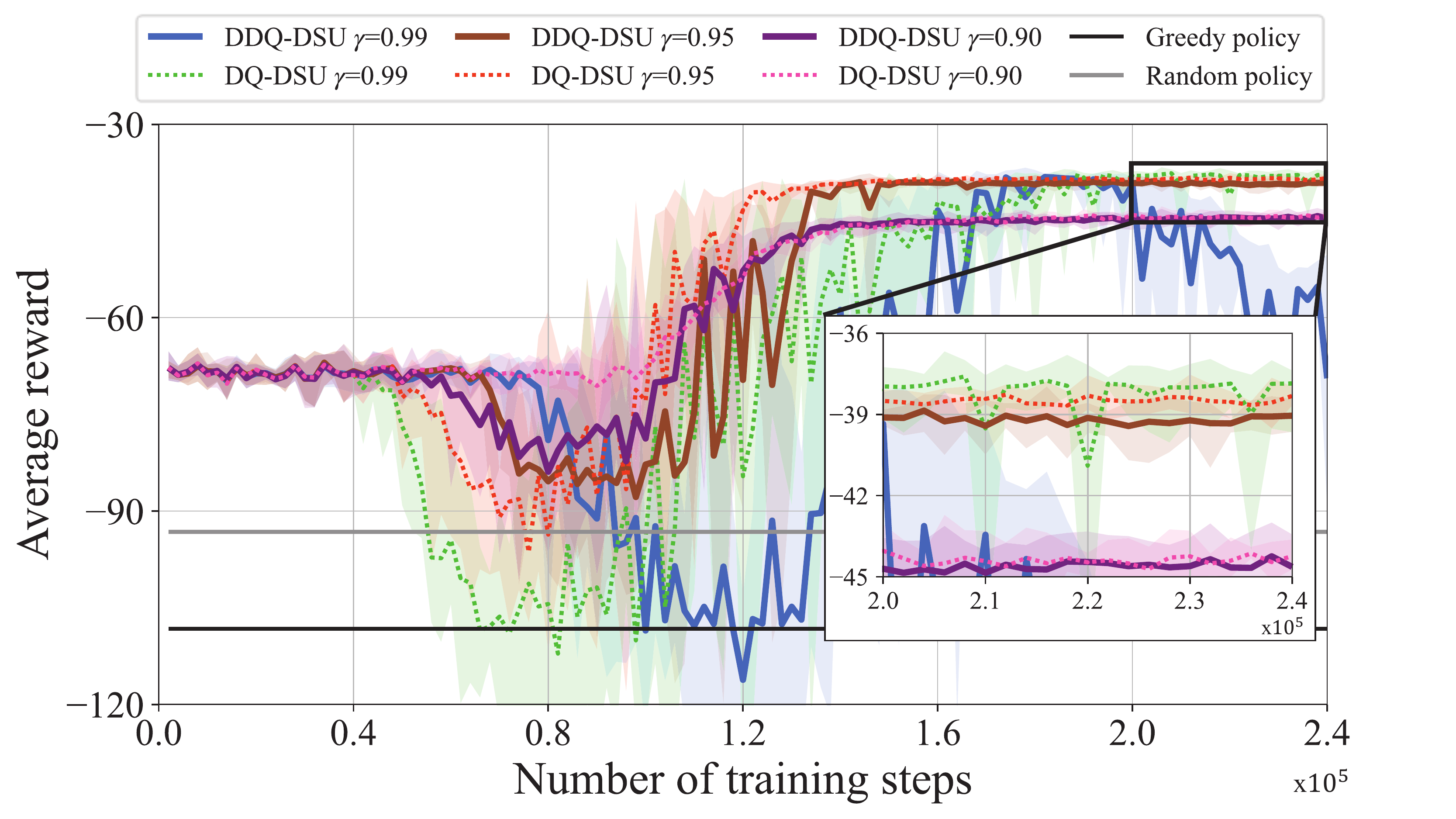}}
\centering \caption{Convergence comparison of: (a) our proposed DDR-DSU and baseline DRL algorithms; (b) baseline policies and DRL algorithms with different discount factor $\gamma$, where $N=40$ and $\beta_1 = \beta_2 =1$. Here, training refers to the operation of replaying and updating.} \label{Fig:Convergence_40User}
\end{figure}

Particularly, we demonstrate the convergence comparison with $N=24$ and $N=40$ users in Figs. \ref{Fig:Convergence_24User} and \ref{Fig:Convergence_40User}, respectively. From these figures, several interesting observations are due. First, for our proposed DRN based algorithms, i.e., the DDR-DSU and DR-DSU, the convergence can be steadily achieved and, after the convergence is attained, the corresponding performance, in terms of the achieved average reward, is superior to those of DQN based algorithms, i.e., the DDQ-DSU and DQ-DSU. Besides, by incorporating the dueling architecture in the DRN based algorithm, the convergence is speeded up. For instance, when there are $N=24$ users (i.e., Fig. \ref{Fig:Convergence_24User} (a)) DDR-DSU and DR-DSU converge after about $1.36*10^5$ and $1.52*10^5$ training steps, respectively. Second, for DQN based algorithms, to achieve a better performance, the discount factor $\gamma$ should be appropriately adjusted: A larger $\gamma$ (e.g., 0.99) is likely to lead to instabilities, while a smaller $\gamma$ (e.g., 0.9) may result in a degraded average reward. To further verify this, we present the mean and standard deviation of the achieved average reward during the last 10 (i.e., 111-120) evaluations in Table \ref{Tab:Mean_SD_Comp}, where the best results are marked in bold. Last but not least, the DRN based algorithms scale well when the size of state-action pairs becomes larger. For instance, when the number of users grows from 24 to 40, the total number of state-action pairs soars by about $160^{(41-25)}\approx1.84*10^{35}$ times, while to reach the convergence, the number of required training steps for DDR-DSU and DR-DSU merely increase about $4.41\%$ (i.e., from $1.36*10^5$ to $1.42*10^5$) and $3.95\%$ (i.e., from $1.52*10^5$ to $1.58*10^5$), respectively.

\begin{table}[!t]
\renewcommand{\arraystretch}{1.3}
\caption{Mean and standard deviation of average reward achieved by DRL algorithms during the last 10 evaluations.}
\label{Tab:Mean_SD_Comp}
\centering
\begin{tabular}{| l |c| c| c| c|}
\hline
\multirow{2}{*}{\bfseries DRL Algorithm} & \multicolumn{2}{c|}{\bfseries  Mean}  & \multicolumn{2}{c|}{\bfseries Standard Deviation} \\
\cline{2-3} \cline{4-5}
{} & {\bfseries $N=24$} & {\bfseries $N=40$} & {\bfseries $N=24$} & {\bfseries $N=40$}            \\
\hline
DDR-DSU                                           & \textbf{-36.59}     &-36.86       & 0.19        & 0.23
\\ \hline
DR-DSU                                            & -36.67      &  \textbf{-36.73}        &\textbf{0.17}        & \textbf{0.17}
\\ \hline
DDQ-DSU, $\gamma = 0.9$                           & -43.27      & -44.55        & 0.41        & 0.78
\\ \hline
DQ-DSU, $\gamma = 0.9$                            & -44.19      & -44.39        & 0.54        & 0.55
\\ \hline
DDQ-DSU, $\gamma = 0.95$                          & -38.38      & -39.24        & 0.34        & 0.56
\\ \hline
DQ-DSU, $\gamma = 0.95$                           & -38.35      & -38.48        & 0.30        & 0.39
\\ \hline
DDQ-DSU, $\gamma = 0.99$                          & -43.57      & -60.69        & 10.35        & 13.41
\\ \hline
DQ-DSU, $\gamma = 0.99$                           & -39.23      & -38.06        & 4.69        & 1.06
\\
\hline
\end{tabular}
\end{table}

\subsection{Effectiveness evaluation}
In this subsection, we evaluate the performance of our proposed algorithm DDR-DSU in different scenarios where the number of users $N$, the request probability $P_n$, and the weight parameter $\beta_2$ are set to different values. As presented in Section \ref{Sec:Simulation}-B, when the convergence is attained, the performance of the algorithm with and without the dueling structure, e.g., DDR-DSU and DR-DSU, are roughly the same. Meanwhile, for the DQN based algorithms, the discount factor $\gamma$ is not recommended to be set as $0.99$, since it is more likely to lead to instabilities. As such, for clear exposition, we only compare the performance of DDR-DSU and its DQN based version DDQ-DSU with $\gamma$ set to $0.9$ and $0.95$ and, as in Section \ref{Sec:Simulation}-B, demonstrate the mean and standard deviation of the average reward achieved over the last 10 evaluations in Figs. \ref{Fig:Perform_Com_N}-\ref{Fig:Perform_Com_B2}. All simulation results are obtained by averaging over 6 independent runs and the same seed is adopted for all algorithms in one run.

\begin{figure} [!t]
\centering
\leavevmode \epsfxsize=3.0in  \epsfbox{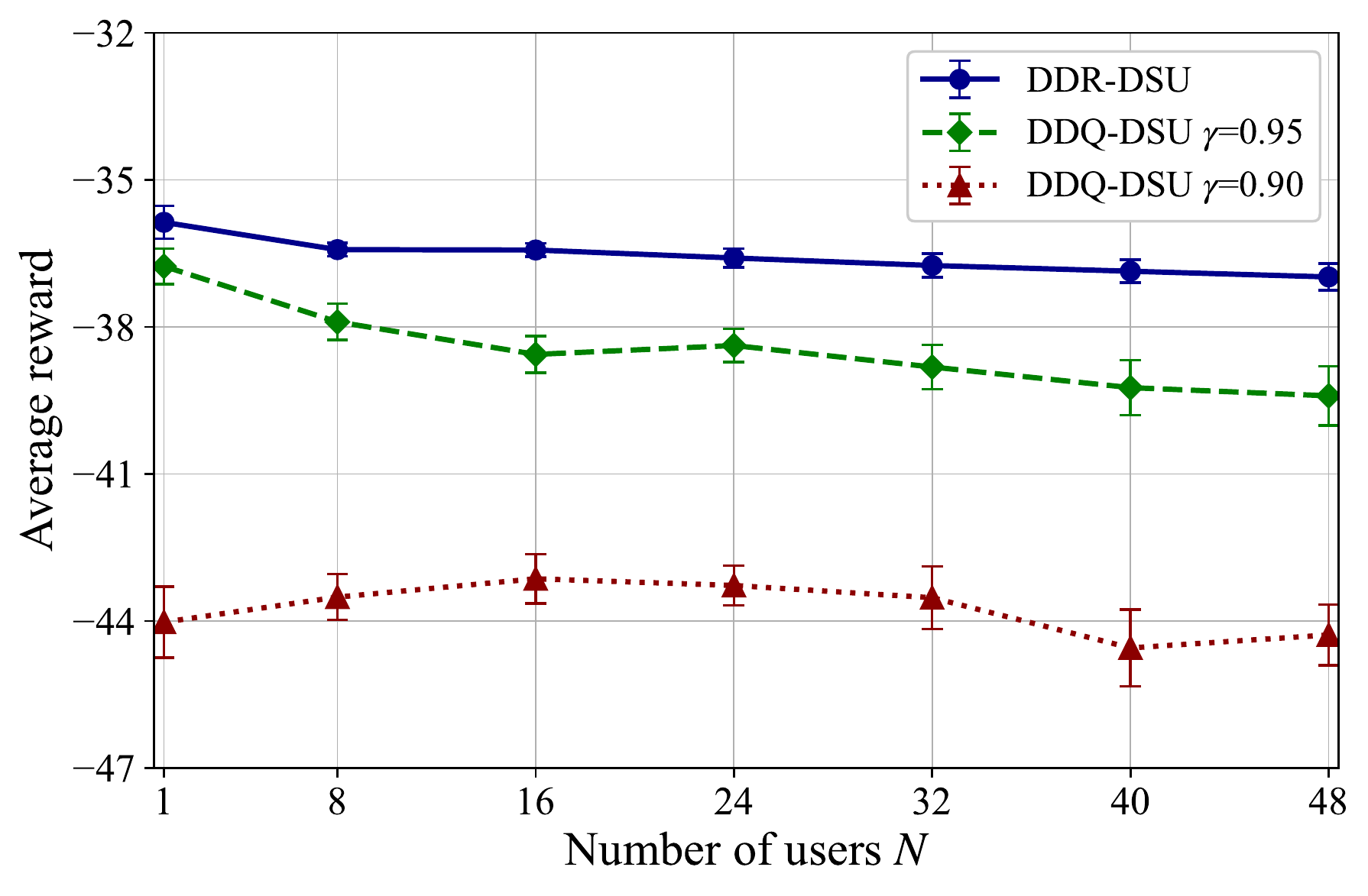}
\centering \caption{Performance comparison in terms of the achieved long-term average reward, where the number of users $N$ varies from 1 to 48 with $P_n = 0.6$ and $\beta_1=\beta_2=1$.} \label{Fig:Perform_Com_N}
\end{figure}

The performance comparison, in terms of achieved average reward, of our devised DDR-DSU and DDQ-DSU is illustrated in Fig. \ref{Fig:Perform_Com_N}, where the number of users $N$ increases from 1 to 48. We note that, as presented in Section \ref{Sec:Simulation}-B, although the size of state-action pairs increases exponentially with respect to the number of users $N$, the performance of DDR-DSU is the best and keeps stable with respect to $N$. In other words, DDR-DSU is able to attain a high average reward while maintaining a low standard deviation.
For instance, when $N=48$, the obtained average reward of DDR-DSU, DDQ-DSU with $\gamma = 0.9$, and DDQ-DSU with $\gamma = 0.95$ are -36.98, -44.28 and -39.41, respectively, and meanwhile, the three corresponding standard deviations are 0.27, 0.62 and 0.61. During the simulation, the average rewards achieved by the random and greedy policies are also calculated, which are not sensitive\footnote{The reason for this non-sensitivity is mainly due to the fact that, as shown in (\ref{Eq:AoI_Cost_Per_Slot})-(\ref{Eq:Totoal_Cost}), the instantaneous cost (i.e., the negative of the instantaneous reward) is not sensitive to the number of users since the total AoI related cost is already averaged over $N$.} to $N$ and keep about -93 and -108, respectively. As such, we remove them from the figure to clearly show the superiority of DDR-DSU. Besides, it is interesting to see that the average reward achieved by our algorithm slightly decreases with respect to the number of users $N$. This is mainly due to the fact that, with the given request probability, when there are less users, the probability of no user request arrival in one time step increases, and there would be a higher probability that no DDP is incorporated in one time step. Accordingly, on the one hand, for a given status update policy, the achieved average AoI may be lowered, bringing a higher average reward. On the other hand, with the same network operation time, there would be more decision epoches for the ECN to update cached data packets to further improve the average reward. To verify this, we have calculated the average AoI related cost (AA) and average energy consumption related cost (AE) for our proposed algorithm and baseline algorithms/policies, and present the results in Fig. \ref{Fig:Perform_Com_N_Bar}.\footnote{In Fig. \ref{Fig:Perform_Com_N_Bar} (b), the y-axis is partially omitted to clearly show the superiority of DDR-DSU. This method is also adopted in following Figs. 9, 11, and 12.} It can be seen from Fig. \ref{Fig:Perform_Com_N_Bar} that for smaller $N$, our proposed algorithm DDR-DSU could bring a higher average reward by lowering the AA with a higher AE.

\begin{figure} [!t]
\centering
\subfigure[] {\leavevmode \epsfxsize=3.0in  \epsfbox{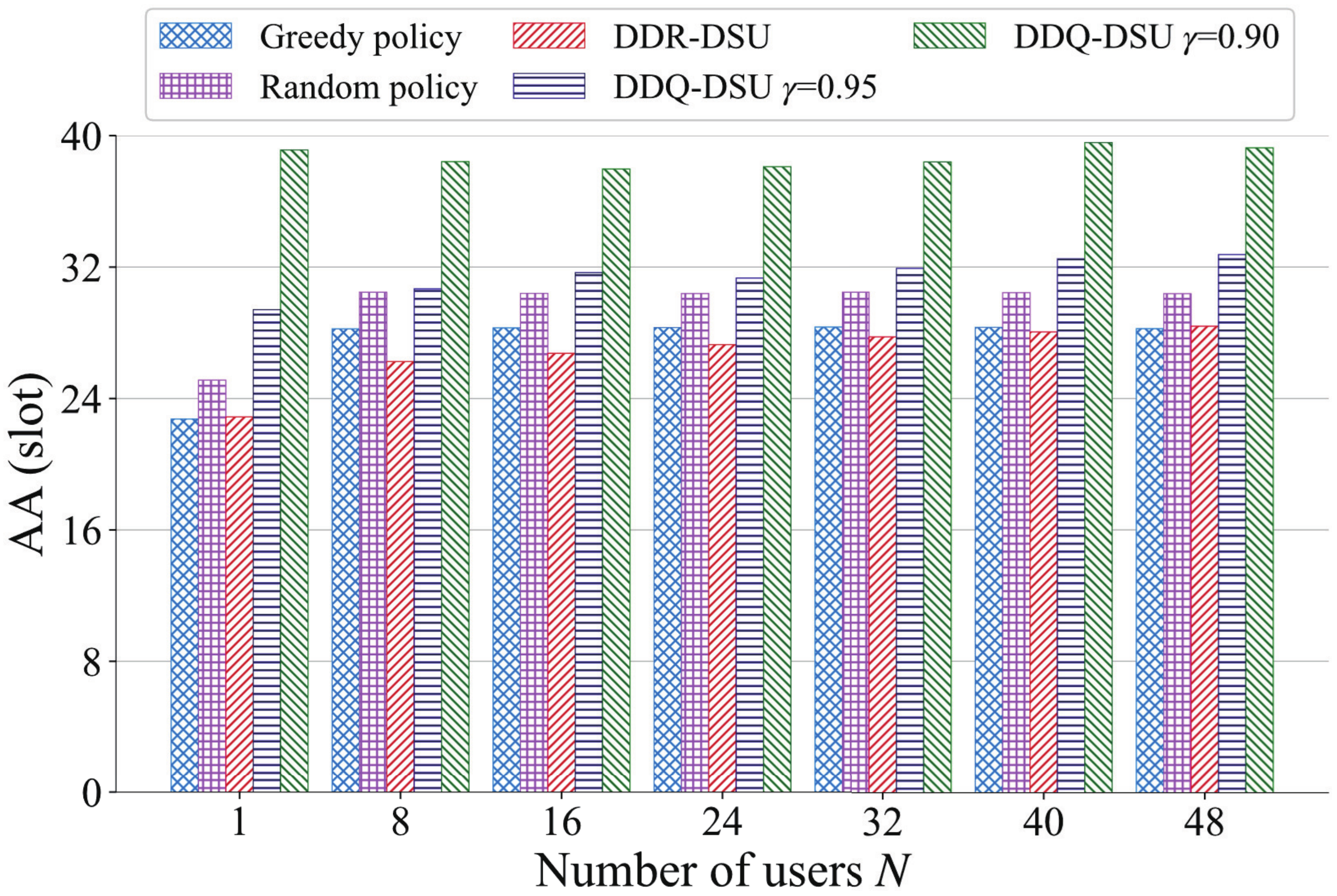}}
\subfigure[] {\leavevmode \epsfxsize=3.0in  \epsfbox{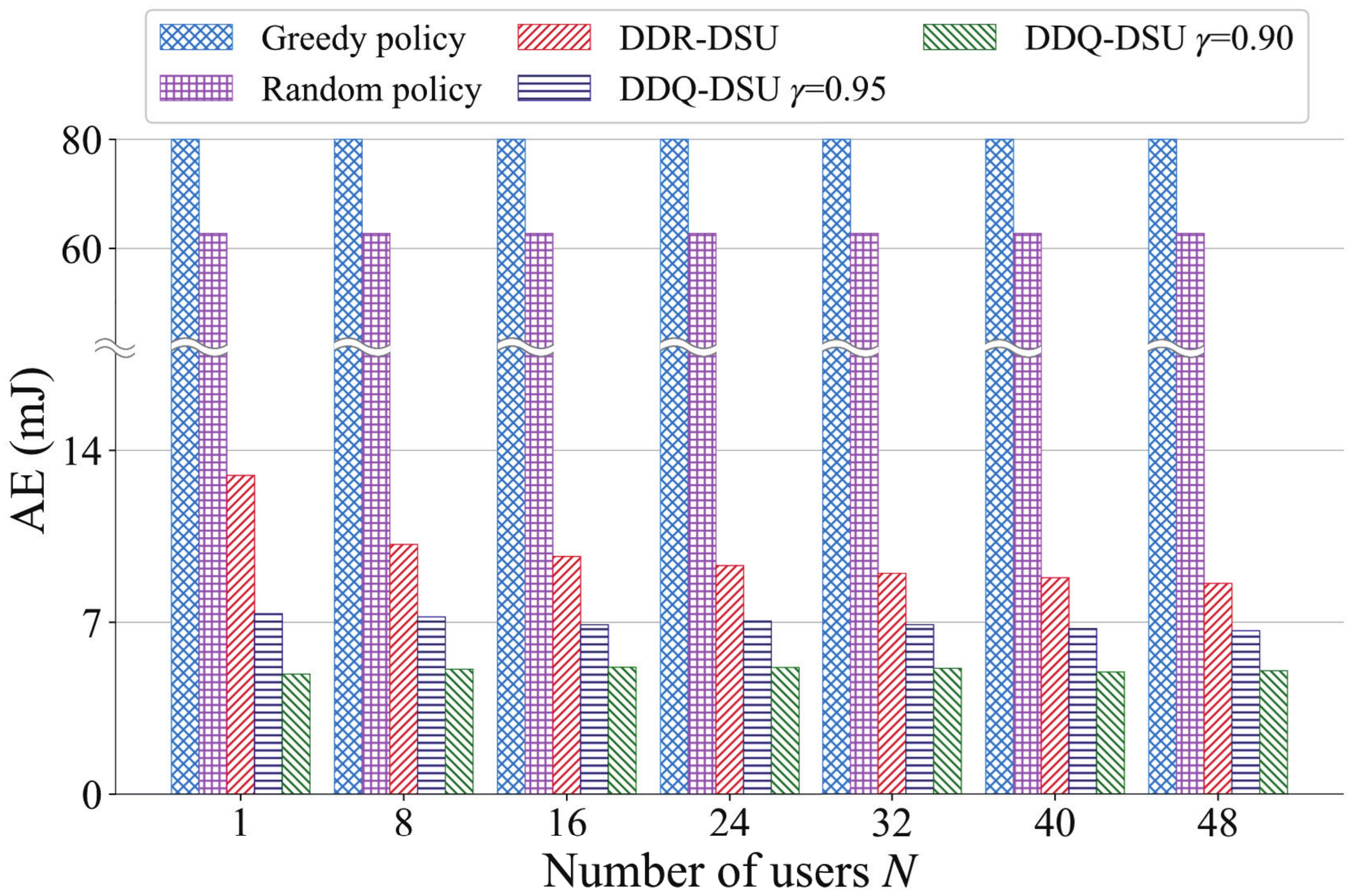}}
\centering \caption{Performance comparison in terms of: (a) AA; (b) AE,  where the number of users $N$ varies from 1 to 48 with $P_n = 0.6$ and $\beta_1=\beta_2=1$.} \label{Fig:Perform_Com_N_Bar}
\end{figure}

\begin{figure} [!t]
\centering
\leavevmode \epsfxsize=3.0in  \epsfbox{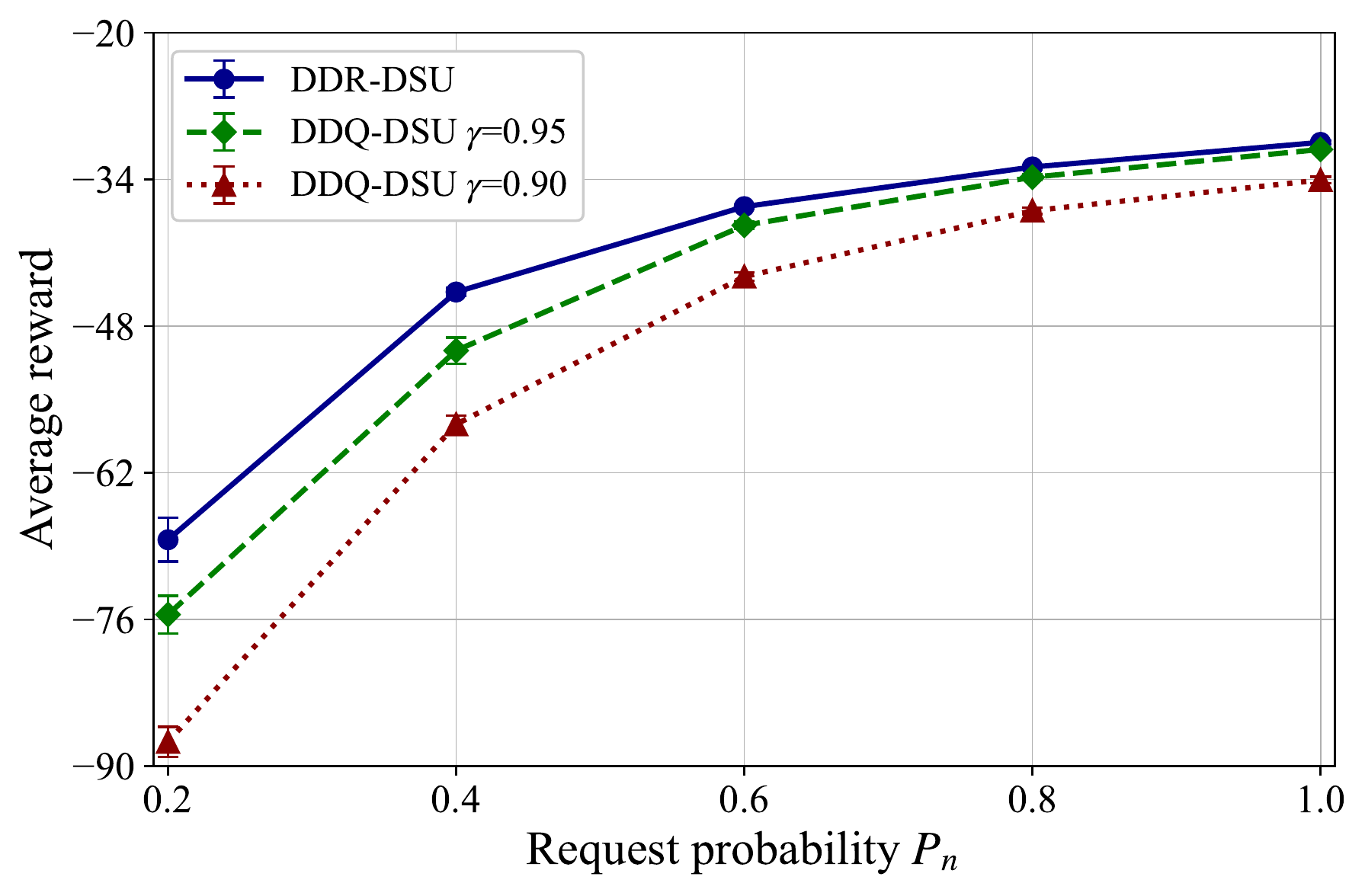}
\centering \caption{Performance comparison in terms of the achieved long-term average reward, where the request probability $P_n$ varies from 0.2 to 1.0 with $N = 24$ and $\beta_1=\beta_2=1$.} \label{Fig:Perform_Com_Pn}
\end{figure}

Fig. \ref{Fig:Perform_Com_Pn} shows the average reward achieved by different algorithms when the request probability $P_n$ varies from 0.2 to 1.0. It can be observed that, for all algorithms, the achieved average reward increases with respect to $P_n$. This is mainly because that, as shown in (\ref{Eq:AoI_Evolution_Per_User_Brief}), the AoI experienced by one user can be reduced only when a fresher update packet is obtained, the probability of which increases with $P_n$. Besides, as shown in this figure, the proposed DDR-DSU always outperforms DDQ-DSU with $\gamma = 0.95$, while the gap between them gradually decreases with respect to $P_n$. This hinges on the fact that when $P_n$ approaches to 1, the average AoI achieved by DDR-DSU is asymptotically lower bounded by that of the greedy policy. In fact, when $P_n =1$ the greedy policy does bring the smallest average AoI value to all users. To verify this, we have calculated the achieved AA and AE of DDR-DSU and baseline algorithms/policies as demonstrated in Fig. \ref{Fig:Perform_Com_Pn_Bar}. It is interesting to see that, in stark contrast to the greedy policy, our proposed algorithm can learn the pattern of user's request and hence, \textit{simultaneously decreases both the average AoI and energy consumption.} For instance, when $P_n$ varies from 0.2 to 0.6, both the achieved AA and AE of DDR-DSU are smaller than those of the greedy policy.

\begin{figure} [!t]
\centering
\subfigure[] {\leavevmode \epsfxsize=3.0in  \epsfbox{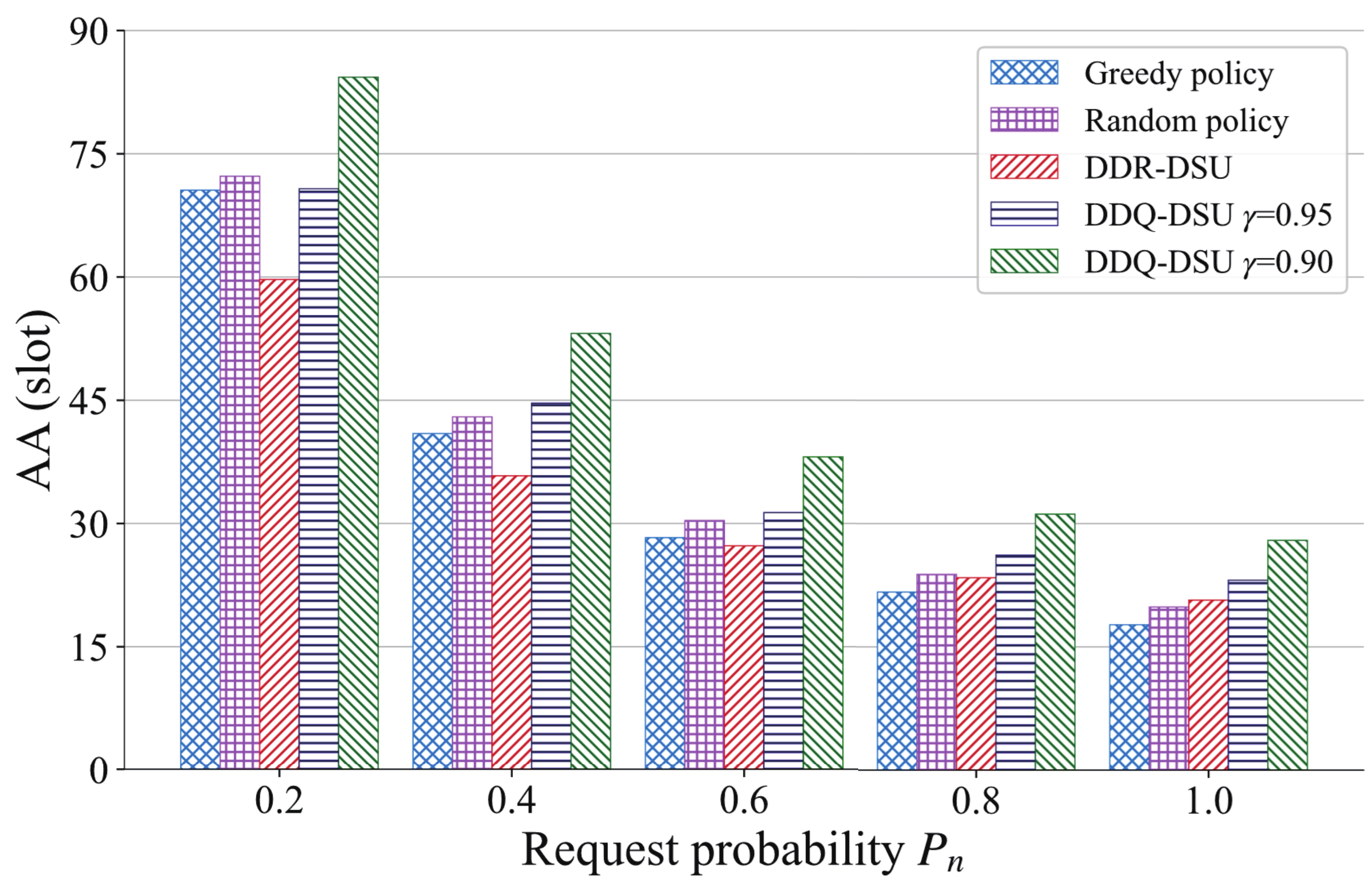}}
\subfigure[] {\leavevmode \epsfxsize=3.0in  \epsfbox{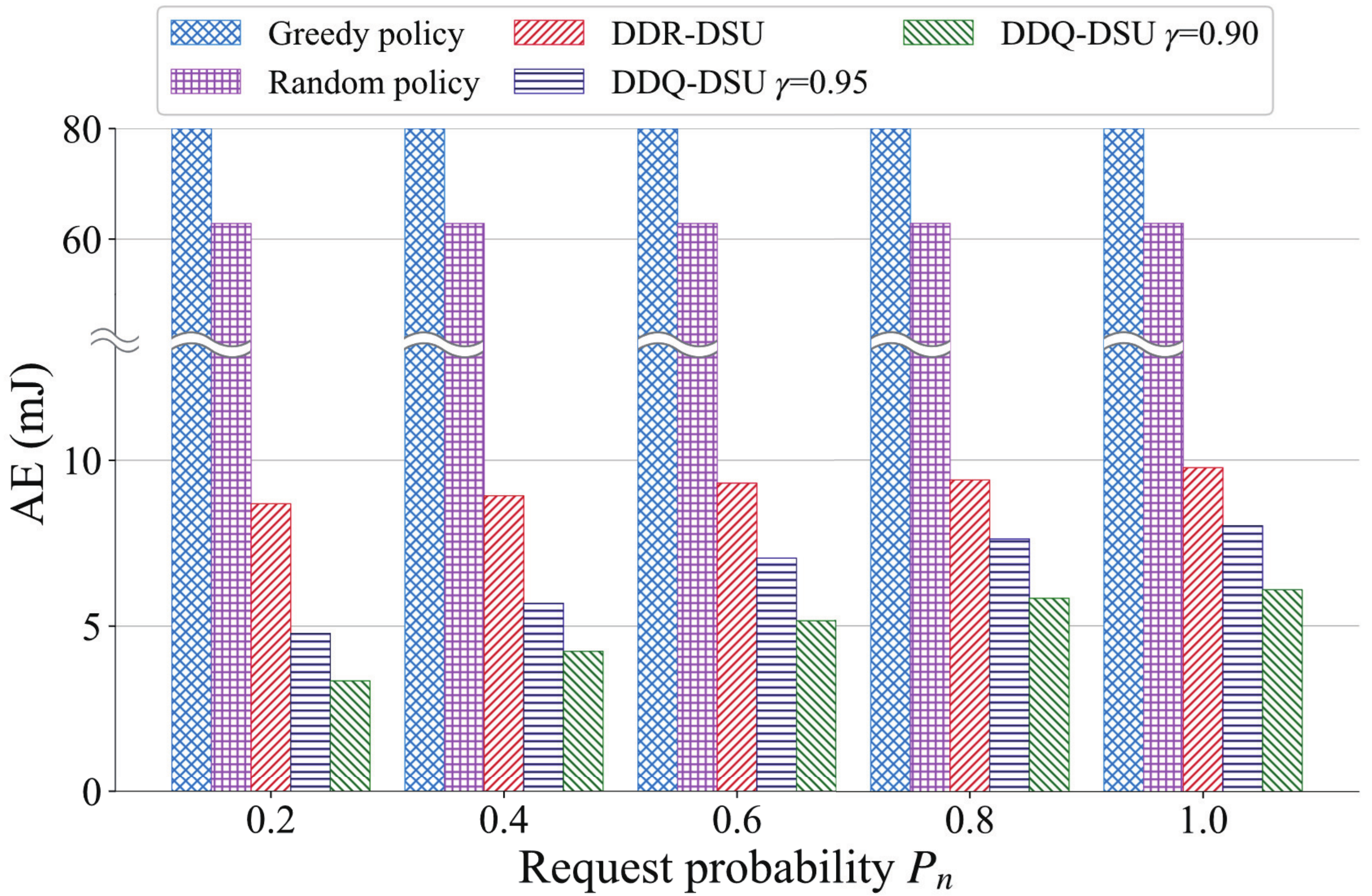}}
\centering \caption{Performance comparison in terms of: (a) AA; (b) AE,  where the request probability $P_n$ varies from 0.2 to 1.0 with $N = 24$ and $\beta_1=\beta_2=1$.} \label{Fig:Perform_Com_Pn_Bar}
\end{figure}

\begin{figure} [!t]
\centering
\leavevmode \epsfxsize=3.0in  \epsfbox{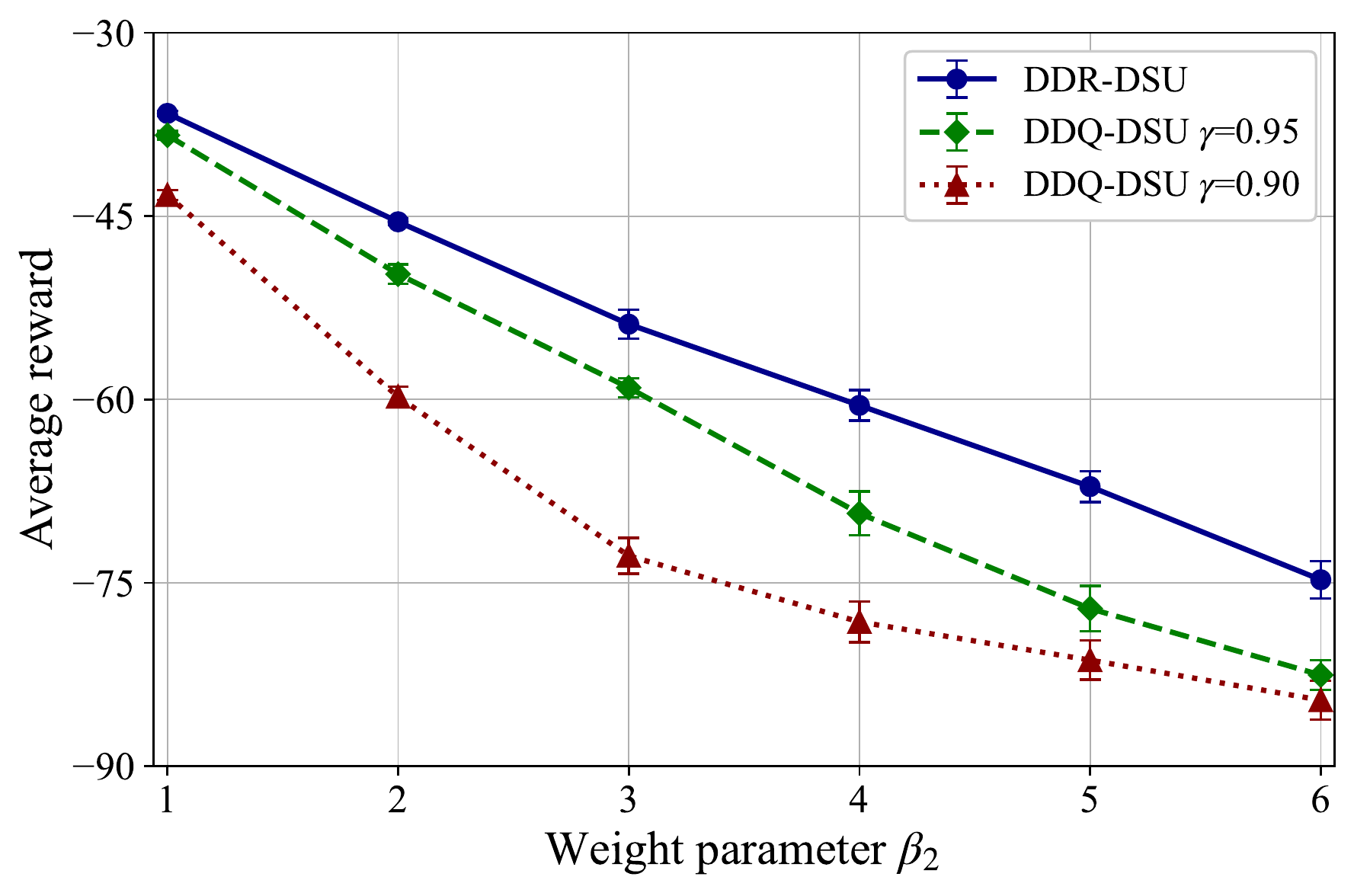}
\centering \caption{Performance comparison in terms of the achieved long-term average reward, where the weight parameter $\beta_2$ varies from 1 to 6 with $N=24$ and $P_n = 0.6$.} \label{Fig:Perform_Com_B2}
\end{figure}

In Fig. \ref{Fig:Perform_Com_B2} we present the simulation results by keeping the weight parameter $\beta_1 = 1$, while changing $\beta_2$ from 1 to 6, which, as shown in  (\ref{Eq:Totoal_Cost}), means decreasing the energy consumption gradually becomes the main concern. Besides, auxiliary to Fig. \ref{Fig:Perform_Com_B2}, we separately show the obtained AA and AE by implementing our algorithm DDR-DSU and baseline algorithms/policies in Fig. \ref{Fig:Perform_Com_B2_Bar} to present more information. As shown in Fig. \ref{Fig:Perform_Com_B2} as well as previous figures, our proposed DDR-DSU always outperforms the DQN based algorithms. Additionally, it can be observed from Figs. \ref{Fig:Perform_Com_B2} and \ref{Fig:Perform_Com_B2_Bar} that, DDR-DSU could well strike the balance between the average AoI experienced by users and energy consumed by sensors as needed and hence, always achieves the largest average reward.

\begin{figure} [!t]
\centering
\subfigure[] {\leavevmode \epsfxsize=3.0in  \epsfbox{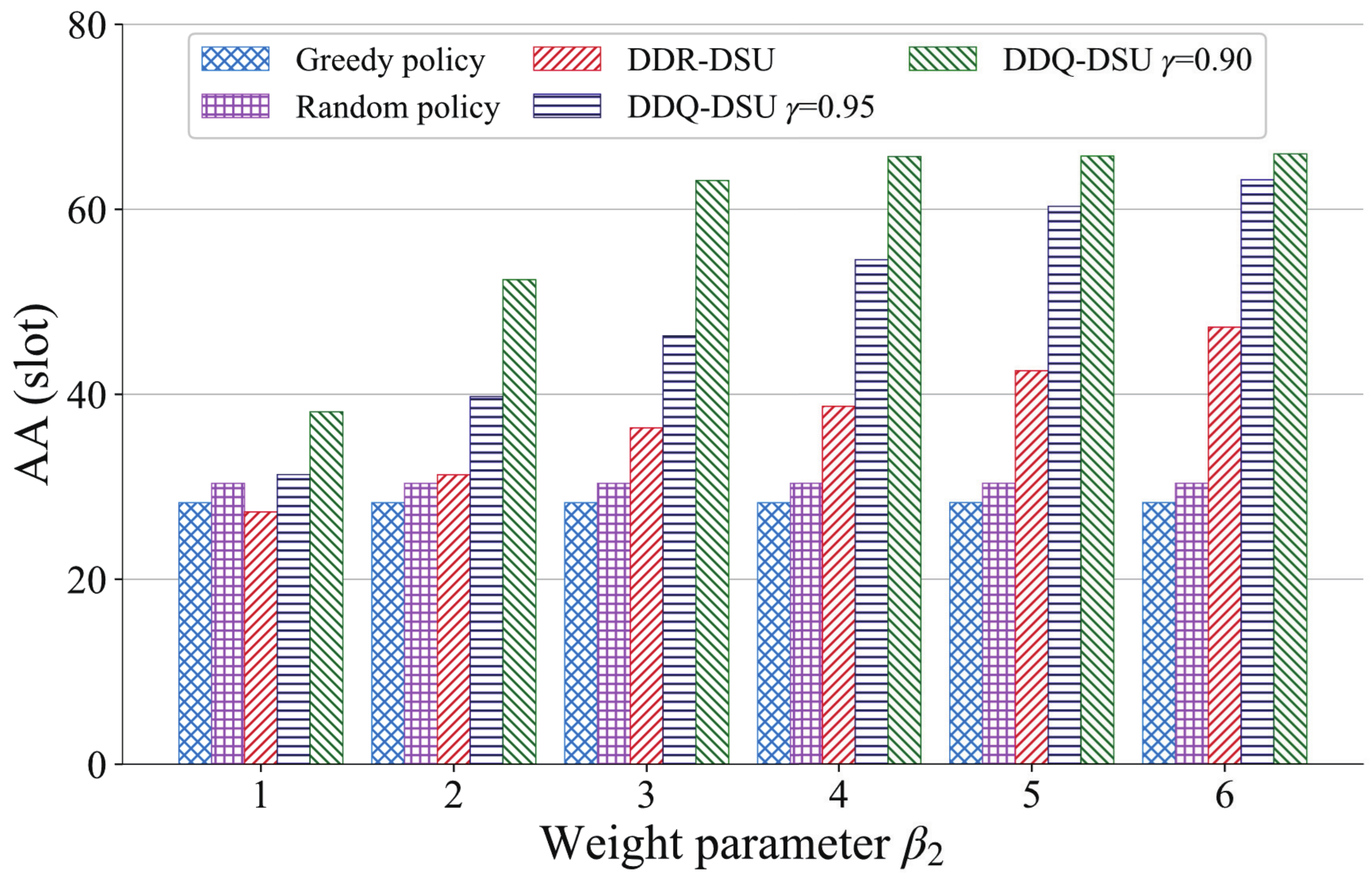}}
\subfigure[] {\leavevmode \epsfxsize=3.0in  \epsfbox{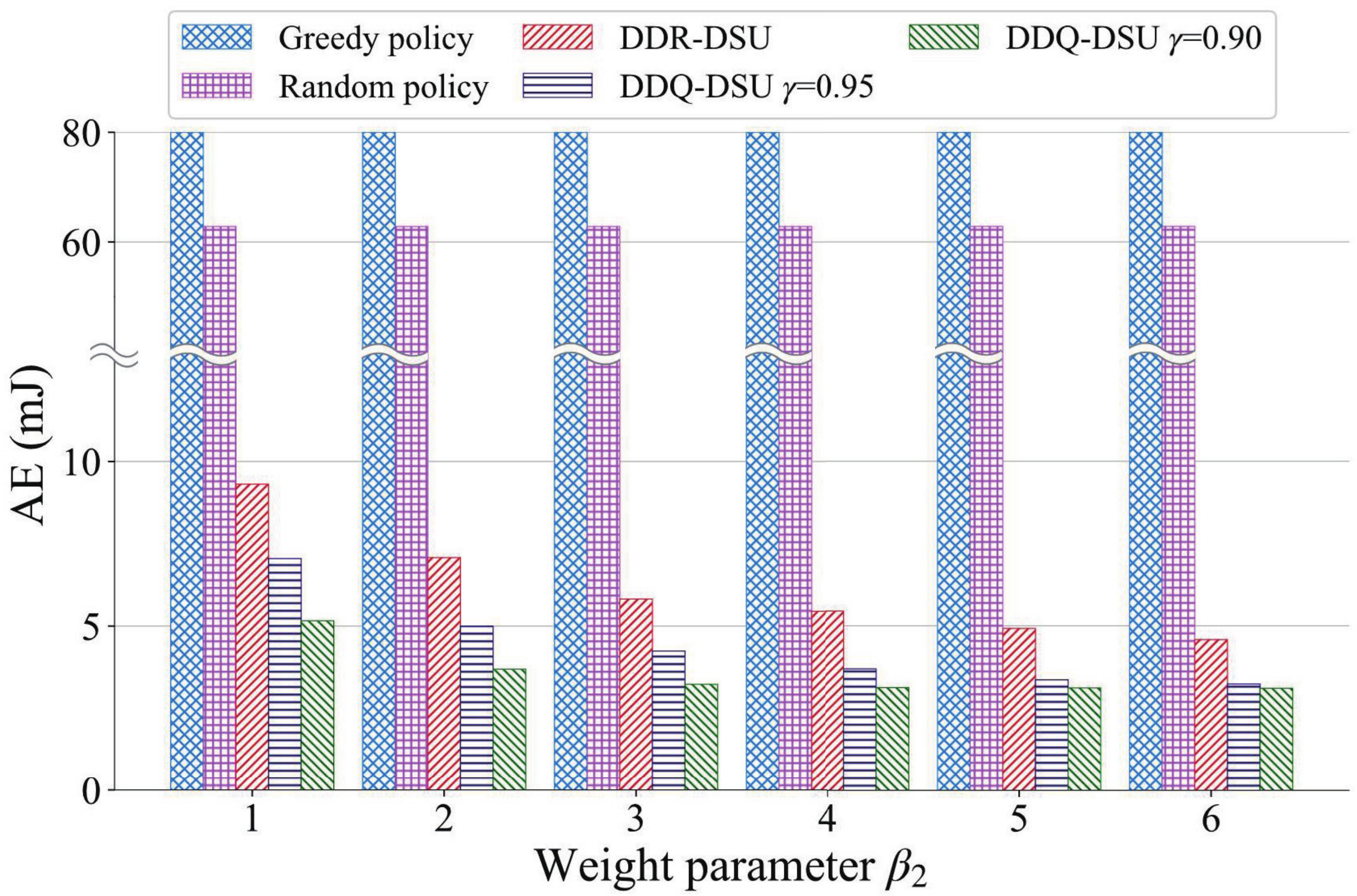}}
\centering \caption{Performance comparison in terms of: (a) AA; (b) AE, where the weight parameter $\beta_2$ varies from 1 to 6 with $N=24$ and $P_n = 0.6$.} \label{Fig:Perform_Com_B2_Bar}
\end{figure}

Finally, we evaluate the performance of our algorithm in the scenario with heterogeneous users and sensors. Particularly, we divide the $N=24$ users into 4 groups, each of which has 6 homogeneous users in terms of the request probability and the associated content preference. Moreover, we divide the $K=8$ sensors into 4 groups according to their transmission failure probabilities, as specified in Section IV-A, i.e., 2 sensors in each group. For the $i$-th user group, $\forall i \in \{1, 2, 3, 4\}$, each user's request probability is set as $0.2*i$, and her preference to the sensor belonging to the group with the same index $i$ is set to be $0.35$, while that value to each sensor in other groups is $0.05$. As such, the sensors in group 1 and group 4 respectively have the lowest and highest peculiarity. We have calculated the average AoI experienced by users in each user group and the average energy consumption of sensors in each sensor group, and present simulation results in Fig. \ref{Fig:Hete_User_Sensor}.

\begin{figure} [!t]
\centering
\subfigure[] {\leavevmode \epsfxsize=3.0in  \epsfbox{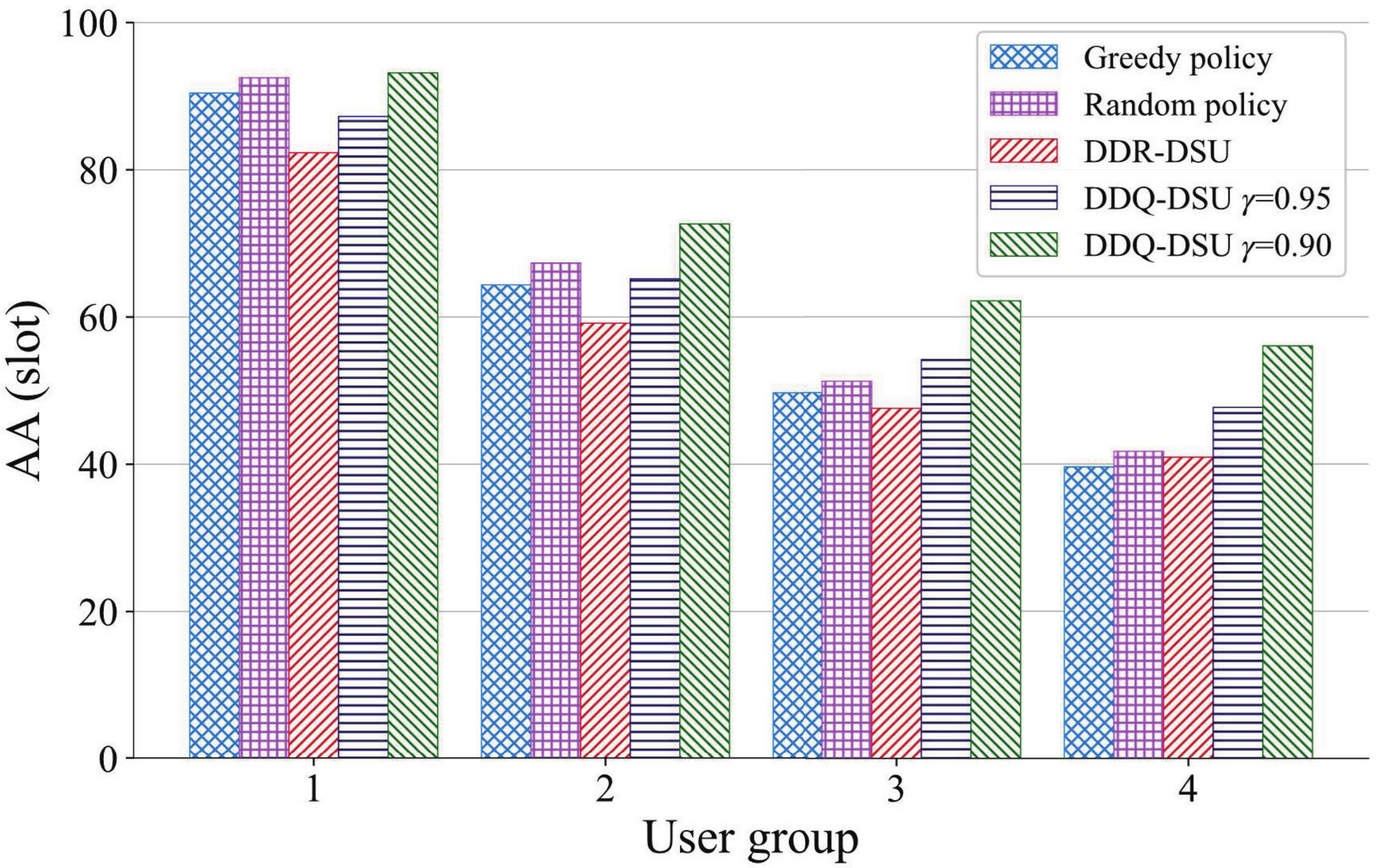}}
\subfigure[] {\leavevmode \epsfxsize=3.0in  \epsfbox{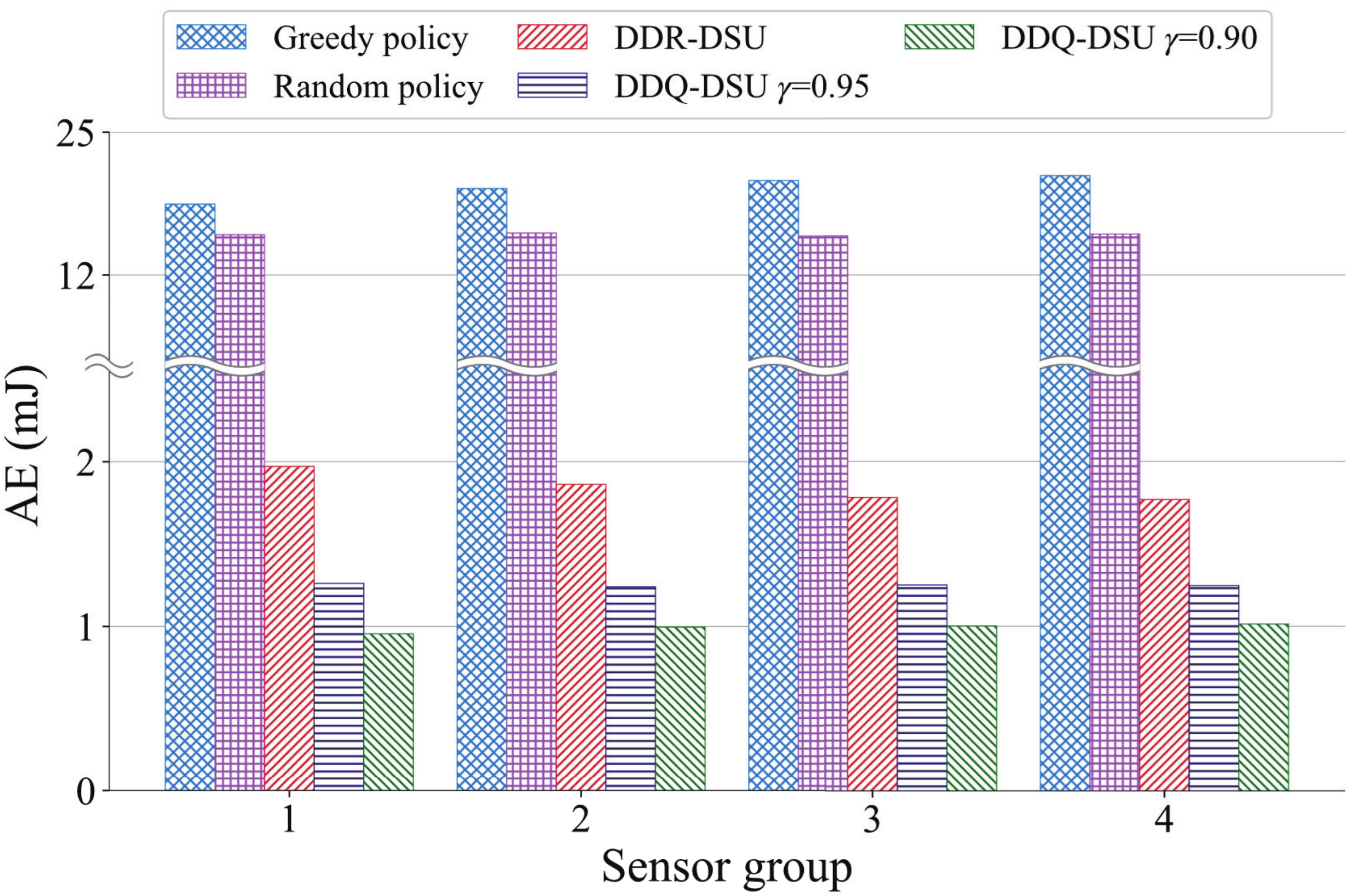}}
\centering \caption{Performance comparison in terms of: (a) AA in different user groups; (b) AE in different sensor groups.}
\label{Fig:Hete_User_Sensor}
\end{figure}

It can be seen from Fig. \ref{Fig:Hete_User_Sensor} that, compared with the greedy policy, our proposed algorithm can adaptively decrease the average AoI experienced by heterogeneous users and the energy consumed by heterogeneous sensors, which is similar to those presented in previous figures. Besides, two interesting observations are made on the performance of our proposed algorithm DDR-DSU. First, as shown in Fig. \ref{Fig:Hete_User_Sensor} (a), users in the group with a smaller index experience a higher average AoI, since they have a smaller request probability and suffer from longer synchronization intervals. Second, as demonstrated in Fig. \ref{Fig:Hete_User_Sensor} (b), the average energy consumption of sensors decreases with respect to the sensor-group index. This can be ascribed to the fact that, to avoid the user with a smaller request probability (i.e., user in the group with a smaller index) from experiencing high AoI, their interested data packets need to be updated more often. Meanwhile, for the concerned simulation scenario, the status updates generated by sensors in the $i$-th group are just the most popular data to the $i$-th user-group. To this end, the sensors in the group with a smaller index are asked to update their status more often, thereby consuming more energy.

\subsection{Execution efficiency and hardware requirement}
At last, we evaluate the computational efficiency and hardware requirement of our proposed algorithm DDR-DSU in simulations. Particularly, for the scenarios with $N=24$ and $N=48$ users, we have recorded the average execution time for one forward pass (i.e., inference or action generation), one backward pass (i.e., minibatch training), and one simulation (consisting of $2.4*10^5$ training steps and 120 evaluations), and moreover, calculated the number of parameters, FLOPs (Floating Point Operations) for one forward pass, and required storage space of the DL model utilized by DDR-DSU, which are summarized in Table \ref{Tab:Exe_Time_Hardware_Req}.

\begin{table}[!t]
\renewcommand{\arraystretch}{1.3}
\caption{Execution time for one forward, one backward and one simulation consumed by, and the number of parameters, FLOPs and storage space of the DL model utilized by DDR-DSU.}
\label{Tab:Exe_Time_Hardware_Req}
\centering
\begin{tabular}{|c|c|c|c|}
\hline
$N$ & Forward (ms) & Backward (ms) & One simulation (hours) \\ \hline
24         & 1.41         & 13.17         & 1.6                    \\ \hline
48         & 1.41         & 15.10         & 1.8                    \\ \hline
$N$ & Parameters & FLOPs   & Storage space (kB) \\ \hline
24          & 159,139    & 317,603 & 639                   \\ \hline
48          & 208,291    & 415,907 & 831                 \\ \hline
\end{tabular}
\centering
\end{table}

It can be seen from Table \ref{Tab:Exe_Time_Hardware_Req} that the time spent in one inference (action generation) is merely about 1.41 ms in a 48-user network, allowing the system to insert real-time decisions on status update in practice. Besides, as demonstrated in Table \ref{Tab:Exe_Time_Hardware_Req}, the hardware requirement of our proposed algorithm is relatively low, although a more powerful machine is adopted to perform simulations in this work. Actually, compared with the DL models devised for applications in the fields of Computer Vision (CV) or Natural Language Processing (NLP), our proposed DRL algorithm adopts very tiny ANNs and hence is much less computationally intensive and requires much smaller memory footprints. Particularly, for our proposed DDR-DSU algorithm there are only four independent ANNs being utilized, where each ANN only contains two hidden layers, each of which consists of 128 neurons. Accordingly, for the case with $N=48$ users, the adopted DL model only has 208.29 thousand parameters, occupying 831 kB storage space, and needs 415.9 thousand FLOPs to generate an action (i.e., complete one forward pass). In contrast, taking the VGG-16 Model (a popular DL model for image recognition) as an instance, it has totally 138.34 million parameters, taking up more than 500 MB storage space, and needs 30.94 billion FLOPs to classify a single image (i.e., complete one forward pass) \cite{Luo_2017_ICCV}.

\begin{remark}
Recently, edge intelligence, aiming at enabling the network edge to carry out training and inference of AI models, has attracted tremendous attention from industry and academia, which continually speeds up the development of hardware platforms, DL libraries and other promising techniques \cite{EI_2020,Hardware_2020,TensorFlow_Lite}. Along with this trend, our proposed DRL algorithm is expected to better suit the emerging IoT networks.
\end{remark}

\section{Conclusions}
This work considered a caching enabled IoT network with non-uniform time steps and focused on striking a balance between the AoI experienced by users and energy consumed by sensors. We formulated a non-uniform time step based dynamic status update optimization problem to minimize the long-term average cost. Further, by leveraging the dueling DQN and R-learning, we proposed a dueling deep R-network-based algorithm, termed DDR-DSU, to solve it. Extensive simulation results showed that our developed DDR-DSU outperformed all the baseline DRL algorithms. In contrast, for DQN based algorithms, the performance was dramatically affected by the adopted discount factor, which was, however, non-trivial to be optimally tuned. Additionally, by learning the dynamics of the environment and making well-informed decisions, our proposed algorithm simultaneously achieved both the lower average AoI and energy consumption than the greedy policy. Based on this work, one interesting extension is to consider the scenario where there are correlations among the status updates from different sensors. Then, to design an efficient status update strategy for sensors, it is essential to consider their sensing correlations and transmission interactions simultaneously.



\appendices
\section{Deep R-network based dynamic status update (DR-DSU) algorithm} \label{Append:DR_DDU}

In contrast to the DDR-DSU algorithm shown in Fig. \ref{Fig:Fig2}, there is only one stream in DR-DSU algorithm, with which the action-value function is approximated. Therefore, the deep R-network (DRN) and target deep R-network (TDRN) are respectively parameterized by two sets of parameters, i.e., $\mathbf \theta$ and  $\mathbf \theta^-$. Accordingly, some equations utilized in Algorithm \ref{Alg:DDR_DDU_algorithm} should be revised for the DR-DSU algorithm, which are shown as follows
\begin{align} \label{Eq:Target_Calculate_Rev}
\hat R_i = U_i- \bar U + \mathop {\max }\limits_{\mathbf A' \in \mathbb A} {R}\left(\mathcal S_i',  \mathbf A'; \mathbf {\theta^-}\right)
\end{align}
\begin{align}   \label{Eq:TD_Error_Rev}
\delta_i =  \hat R_i - {R}\left(\mathcal S_i,  \mathbf A; \mathbf {\theta}\right)
\end{align}
and
\begin{align}   \label{Eq:Net_Par_Updating_Rev}
\mathbf \theta = \mathbf \theta - \alpha \nabla_{\theta} L\left( \mathbf {\theta} \right),\ {\rm{with}} \ L\left( \mathbf {\theta} \right) = \frac{1}{D_b} \sum\nolimits_{i =1}^{D_b} (\delta_i) ^2.
\end{align}
The pseudo-code of DR-DSU is presented in Algorithm \ref{Alg:DR_DDU_algorithm}.

\begin{algorithm}[!t]
\caption{Deep R-network based dynamic status update (DR-DSU) algorithm.}
\begin{algorithmic}[1] \label{Alg:DR_DDU_algorithm}
\STATE \textbf{Initialization:}
\STATE Initialize the experience replay buffer $\mathbb D$, average reward $\bar U$, DRN parameters $\mathbf {\theta}$, TDRN parameters $\mathbf {\theta^{-}} = \mathbf {\theta}$, maximum number of training
steps $T_{max}$, $t = 1$, and $\mathcal S(t) = \{0, 0, \ldots, 0\}$.
\STATE \textbf{Go into a loop:}
\FOR{$t < T_{\max} $}
\STATE Execute steps \ref{Alg:Repeat_Begin} to \ref{Alg:Repeat_End} in Algorithm \ref{Alg:DDR_DDU_algorithm}.
\IF {$t > T_s$}
\STATE \textbf{Replaying and updating (Training):}
\STATE Uniformly sample a minibatch of $D_b$ tuples from the replay buffer $\mathbb D$ and calculate the TD error for each sampled tuple with (\ref{Eq:Target_Calculate_Rev}) and (\ref{Eq:TD_Error_Rev}).
\STATE Update average reward $\bar U$ with (\ref{Eq:Ave_reward_Update}).
\STATE Perform a gradient descent step on loss utilizing (\ref{Eq:Net_Par_Updating_Rev}).
\IF {$mod(t, T_0) = 0$}
\STATE \textbf{Updating target network:} Set the parameters $\mathbf {\theta^{-}} = \mathbf {\theta}$.
\ENDIF
\ENDIF
\STATE Set $t = t+1$.
\ENDFOR
\STATE \textbf{Output:} Set the parameters $\mathbf {\theta^{-}} = \mathbf {\theta}$ and output TDRN.
\label{Alg1:Iteration_End}
\end{algorithmic}
\end{algorithm}

\bibliographystyle{IEEEtran}
\bibliography{IEEEabrv,RL_UP_Ref}

\begin{thebibliography}{10}
\providecommand{\url}[1]{#1}
\csname url@samestyle\endcsname
\providecommand{\newblock}{\relax}
\providecommand{\bibinfo}[2]{#2}
\providecommand{\BIBentrySTDinterwordspacing}{\spaceskip=0pt\relax}
\providecommand{\BIBentryALTinterwordstretchfactor}{4}
\providecommand{\BIBentryALTinterwordspacing}{\spaceskip=\fontdimen2\font plus
\BIBentryALTinterwordstretchfactor\fontdimen3\font minus
  \fontdimen4\font\relax}
\providecommand{\BIBforeignlanguage}[2]{{%
\expandafter\ifx\csname l@#1\endcsname\relax
\typeout{** WARNING: IEEEtran.bst: No hyphenation pattern has been}%
\typeout{** loaded for the language `#1'. Using the pattern for}%
\typeout{** the default language instead.}%
\else
\language=\csname l@#1\endcsname
\fi
#2}}
\providecommand{\BIBdecl}{\relax}
\BIBdecl

\bibitem{Survey_IoT_Applications_2015}
A.~Al-Fuqaha, M.~Guizani, M.~Mohammadi, M.~Aledhari, and M.~Ayyash, ``{Internet
  of Things}: A survey on enabling technologies, protocols, and applications,''
  \emph{IEEE Commun. Surveys Tuts.}, vol.~17, no.~4, pp. 2347--2376, Jun. 2015.

\bibitem{IoT_Caching_2016_Network}
M.~Amadeo, C.~Campolo, J.~Quevedo, D.~Corujo, A.~Molinaro, A.~Iera, R.~L.
  Aguiar, and A.~V. Vasilakos, ``Information-centric networking for the
  {Internet of Things}: challenges and opportunities,'' \emph{IEEE Network},
  vol.~30, no.~2, pp. 92--100, Mar. 2016.

\bibitem{Caching_IoT_EH_ICC_2016}
D.~Niyato, D.~I. Kim, P.~Wang, and L.~Song, ``A novel caching mechanism for
  {Internet of Things (IoT)} sensing service with energy harvesting,'' in
  \emph{Proc. IEEE ICC}, May 2016, pp. 1--6.

\bibitem{IoT_Caching_2017_RL}
Y.~He, F.~R. Yu, N.~Zhao, V.~C.~M. Leung, and H.~Yin, ``Software-defined
  networks with mobile edge computing and caching for smart cities: A big data
  deep reinforcement learning approach,'' \emph{IEEE Commun. Magazine},
  vol.~55, no.~12, pp. 31--37, Dec. 2017.

\bibitem{Femtocell_Caching_2013}
K.~Shanmugam, N.~Golrezaei, A.~G. Dimakis, A.~F. Molisch, and G.~Caire,
  ``Femtocaching: Wireless content delivery through distributed caching
  helpers,'' \emph{IEEE Trans. Inf. Theory}, vol.~59, no.~12, pp. 8402--8413,
  Dec. 2013.

\bibitem{Content_Caching_2014}
X.~Wang, M.~Chen, T.~Taleb, A.~Ksentini, and V.~C. Leung, ``Cache in the air:
  Exploiting content caching and delivery techniques for {5G} systems,''
  \emph{IEEE Commun. Magazine}, vol.~52, no.~2, pp. 131--139, Feb. 2014.

\bibitem{Our_Caching_Comm_Mag}
M.~Sheng, C.~Xu, J.~Liu, J.~Song, X.~Ma, and J.~Li, ``Enhancement for content
  delivery with proximity communications in caching enabled wireless networks:
  Architecture and challenges,'' \emph{IEEE Commun. Magazine}, vol.~54, no.~8,
  pp. 70--76, Aug. 2016.

\bibitem{Caching_Multicasting_BZhou_2016}
B.~{Zhou}, Y.~{Cui}, and M.~{Tao}, ``Stochastic content-centric multicast
  scheduling for cache-enabled heterogeneous cellular networks,'' \emph{IEEE
  Trans. Wireless Commun.}, vol.~15, no.~9, pp. 6284--6297, Sep. 2016.

\bibitem{Tang_2019}
J.~Tang, T.~Q. Quek, T.-H. Chang, and B.~Shim, ``Systematic resource allocation
  in cloud {RAN} with caching as a service under two timescales,'' \emph{IEEE
  Trans. Commun.}, vol.~67, no.~11, pp. 7755--7770, Nov. 2019.

\bibitem{Update_WN_Lifetime_JSAC_2018}
S.~O. {Somuyiwa}, A.~Gy{\"o}rgy, and D.~G{\"u}nd{\"u}z, ``A
  reinforcement-learning approach to proactive caching in wireless networks,''
  \emph{IEEE J. Sel. Areas Commun.}, vol.~36, no.~6, pp. 1331--1344, Jun. 2018.

\bibitem{AoI_Org_2012}
S.~Kaul, R.~Yates, and M.~Gruteser, ``Real-time status: How often should one
  update?'' in \emph{Proc. IEEE INFOCOM}, Mar. 2012, pp. 2731--2735.

\bibitem{AoI_Survey_2017}
A.~Kosta, N.~Pappas, and V.~Angelakis, ``Age of information: A new concept,
  metric, and tool,'' \emph{Found. Trends Netw.}, vol.~12, no.~3, pp. 162--259,
  2017.

\bibitem{YanArafaQue:20}
H.~H. Yang, A.~Arafa, T.~Q.~S. Quek, and H.~V. Poor, ``Optimizing information
  freshness in wireless networks: A stochastic geometry approach,''
  \emph{{IEEE} Trans. Mobile Comput.}, Feb. 2020, accepted for publication.

\bibitem{Our_IF_2019}
C.~Xu, H.~H. Yang, X.~Wang, and T.~Q. Quek, ``Optimizing information freshness
  in computing enabled {IoT} networks,'' \emph{IEEE Internet Things J.},
  vol.~7, no.~2, pp. 971--985, Feb. 2020.

\bibitem{BZhou_Samp_Up_2019}
B.~{Zhou} and W.~{Saad}, ``Joint status sampling and updating for minimizing
  age of information in the {Internet of Things},'' \emph{IEEE Trans. Commun.},
  vol.~67, no.~11, pp. 7468--7482, Jul. 2019.

\bibitem{BZhou_Non_Uni_Pack_2020}
------, ``Minimum age of information in the {Internet of Things} with
  non-uniform status packet sizes,'' \emph{IEEE Trans. Wireless Commun.},
  vol.~19, no.~3, pp. 1933--1947, Mar. 2020.

\bibitem{Age_Updating_2017}
R.~D. Yates, P.~Ciblat, A.~Yener, and M.~Wigger, ``Age-optimal constrained
  cache updating,'' in \emph{Proc. IEEE ISIT}, Jun. 2017, pp. 141--145.

\bibitem{Refresh_Rate_AoI_2018}
J.~{Zhong}, R.~D. {Yates}, and E.~{Soljanin}, ``Two freshness metrics for local
  cache refresh,'' in \emph{Proc. IEEE ISIT}, Jun. 2018, pp. 1924--1928.

\bibitem{AoI_Cache_Updating_2019}
H.~Tang, P.~Ciblat, J.~Wang, M.~Wigger, and R.~Yates, ``Age of information
  aware cache updating with file-and age-dependent update durations,''
  \emph{arXiv preprint arXiv:1909.05930}, Sep. 2019.

\bibitem{Ave_AoI_EH_Sensor_Pappas}
N.~{Pappas}, Z.~{Chen}, and M.~{Hatami}, ``Average {AoI} of cached status
  updates for a process monitored by an energy harvesting sensor,'' in
  \emph{Proc. Annual Conference on Information Sciences and Systems}, May 2020,
  pp. 1--5.

\bibitem{ILP_Caching_ICC_2020}
G.~{Ahani} and D.~{Yuan}, ``Accounting for information freshness in scheduling
  of content caching,'' in \emph{Proc. IEEE ICC}, Jul. 2020, pp. 1--6.

\bibitem{IF_Caching_2020}
M.~Bastopcu and S.~Ulukus, ``Information freshness in cache updating systems,''
  ArXiv, Apr. 2020, arXiv:2004.09475v1.

\bibitem{AoI_Dealy_Tradeoff_2020}
S.~Zhang, L.~Wang, H.~Luo, X.~Ma, and S.~Zhou, ``{AoI}-delay tradeoff in mobile
  edge caching with freshness-aware content refreshing,'' Feb. 2020,
  arXiv:2002.05868v1.

\bibitem{DRL_Updating_2020}
M.~Ma and V.~W.~S. Wong, ``A deep reinforcement learning approach for dynamic
  contents caching in {HetNets},'' \emph{ArXiv}, vol. abs/2004.07911, Apr.
  2020, accepted by IEEE ICC'20.

\bibitem{IoT_Caching_2020}
X.~Wu, X.~Li, J.~Li, P.~C. Ching, V.~C.~M. Leung, and H.~V. Poor, ``Caching
  transient content for {IoT} sensing: Multi-agent soft actor-critic,''
  \emph{arXiv preprint arXiv:2008.13191}, Aug. 2020.

\bibitem{Non_Uniform_Time_Step_2020}
Y.~{Yu}, S.~C. {Liew}, and T.~{Wang}, ``Non-uniform time-step deep {Q}-network
  for carrier-sense multiple access in heterogeneous wireless networks,''
  \emph{{IEEE} Trans. Mobile Comput.}, Apr. 2020, accepted for publication.

\bibitem{Dueling_DQN_Org}
Z.~Wang, T.~Schaul, M.~Hessel, H.~Van~Hasselt, M.~Lanctot, and N.~De~Freitas,
  ``Dueling network architectures for deep reinforcement learning,'' in
  \emph{Proc. PMLR}, vol.~48, Jun. 2016, pp. 1995--2003.

\bibitem{RL_1993}
A.~Schwartz, ``A reinforcement learning method for maximizing undiscounted
  rewards,'' in \emph{Proc. ICML}, 1993, pp. 298--305.

\bibitem{Our_TR_2021}
C.~Xu, Y.~Xie, X.~Wang, H.~H. Yang, D.~Niyato, and T.~Q.~S. Quek, ``Optimizing
  the long-term average reward for continuing {MDPs}: A technical report,''
  Apr. 2021, arXiv:2104.06139.

\bibitem{CXu_Updating_2020}
C.~{Xu}, X.~{Wang}, H.~H. {Yang}, H.~{Sun}, and T.~Q.~S. {Quek}, ``{AoI} and
  energy consumption oriented dynamic status updating in caching enabled {IoT}
  networks,'' in \emph{Proc. IEEE INFOCOM'20 AoI WKSHP}, Aug. 2020, pp.
  710--715.

\bibitem{RL_Introduction}
R.~S. Sutton and A.~G. Barto, \emph{Reinforcement learning: An
  introduction}.\hskip 1em plus 0.5em minus 0.4em\relax MIT Press, 2018.

\bibitem{DQN_Nature_Letter}
V.~Mnih, K.~Kavukcuoglu, D.~Silver, A.~A. Rusu, J.~Veness, M.~G. Bellemare,
  A.~Graves, M.~Riedmiller, A.~K. Fidjeland, G.~Ostrovski \emph{et~al.},
  ``Human-level control through deep reinforcement learning,'' \emph{Nature},
  vol. 518, no. 7540, pp. 529--533, Feb. 2015.

\bibitem{DQN_Variations_Ranbow}
M.~Hessel, J.~Modayil, H.~Van~Hasselt, T.~Schaul, G.~Ostrovski, W.~Dabney,
  D.~Horgan, and D.~Silver, ``Rainbow: Combining improvements in deep
  reinforcement learning,'' in \emph{Proc. AAAI}, Feb. 2018, pp. 3215--3222.

\bibitem{DRL_Surrvey_Tao}
N.~C. {Luong}, D.~T. {Hoang}, S.~{Gong}, D.~{Niyato}, P.~{Wang}, Y.~{Liang},
  and D.~I. {Kim}, ``Applications of deep reinforcement learning in
  communications and networking: A survey,'' \emph{IEEE Commun. Surveys Tuts.},
  vol.~21, no.~4, pp. 3133--3174, May 2019.

\bibitem{DDQN_2016}
H.~van Hasselt, A.~Guez, and D.~Silver, ``Deep reinforcement learning with
  double {Q}-learning,'' in \emph{Proc. AAAI}, Mar. 2016, pp. 2094--2100.

\bibitem{Priority_2016}
T.~Schaul, J.~Quan, I.~Antonoglou, and D.~Silver, ``Prioritized experience
  replay,'' in \emph{Proc. ICLR}, May 2016, pp. 1--21.

\bibitem{DF_Tuning_1}
S.~F. Abedin, M.~S. Munir, N.~H. Tran, Z.~Han, and C.~S. Hong, ``Data freshness
  and energy-efficient {UAV} navigation optimization: A deep reinforcement
  learning approach,'' Feb. 2020, arXiv:2003.04816v1.

\bibitem{DF_Tuning_2}
L.~{Wang}, H.~{Ye}, L.~{Liang}, and G.~Y. {Li}, ``Learn to compress {CSI} and
  allocate resources in vehicular networks,'' \emph{IEEE Trans. Commun.},
  vol.~68, no.~6, pp. 3640--3653, Jun. 2020.

\bibitem{DF_Tuning_3}
J.~{Tan}, L.~{Zhang}, Y.~{Liang}, and D.~{Niyato}, ``Intelligent sharing for
  {LTE} and {WiFi} systems in unlicensed bands: A deep reinforcement learning
  approach,'' \emph{IEEE Trans. Commun.}, vol.~68, no.~5, pp. 2793--2808, May
  2020.

\bibitem{Introduction_DRL}
\BIBentryALTinterwordspacing
V.~Fran{\c{c}}ois{-}Lavet, P.~Henderson, R.~Islam, M.~G. Bellemare, and
  J.~Pineau, ``An introduction to deep reinforcement learning,''
  \emph{Foundations and Trends in Machine Learning}, vol.~11, no. 3-4, pp.
  219--354, Dec. 2018. [Online]. Available:
  \url{http://arxiv.org/abs/1811.12560}
\BIBentrySTDinterwordspacing

\bibitem{Wu_2016_CVPR}
J.~Wu, C.~Leng, Y.~Wang, Q.~Hu, and J.~Cheng, ``Quantized convolutional neural
  networks for mobile devices,'' in \emph{Proc. IEEE CVPR}, Jun. 2016.

\bibitem{Complexity_Ana_UAV}
F.~Wu, H.~Zhang, J.~Wu, L.~Song, Z.~Han, and H.~V. Poor, ``{UAV}-to-device
  underlay communications: Age of information minimization by multi-agent deep
  reinforcement learning,'' Mar. 2020.

\bibitem{EC_Sensing}
S.~{Maleki}, A.~{Pandharipande}, and G.~{Leus}, ``Energy-efficient distributed
  spectrum sensing for cognitive sensor networks,'' \emph{IEEE Sensors J.},
  vol.~11, no.~3, pp. 565--573, Mar. 2011.

\bibitem{Adam_Ref}
D.~P. Kingma and J.~Ba, ``Adam: A method for stochastic optimization,'' in
  \emph{Proc. ICLR}, May 2015.

\bibitem{He_Ref}
K.~{He}, X.~{Zhang}, S.~{Ren}, and J.~{Sun}, ``Delving deep into rectifiers:
  Surpassing human-level performance on {ImageNet} classification,'' in
  \emph{Proc. ICCV}, Dec. 2015, pp. 1026--1034.

\bibitem{Luo_2017_ICCV}
J.-H. Luo, J.~Wu, and W.~Lin, ``{ThiNet}: A filter level pruning method for
  deep neural network compression,'' in \emph{Proc. IEEE ICCV}, Oct. 2017.

\bibitem{EI_2020}
X.~{Wang}, Y.~{Han}, V.~C.~M. {Leung}, D.~{Niyato}, X.~{Yan}, and X.~{Chen},
  ``Convergence of edge computing and deep learning: A comprehensive survey,''
  \emph{IEEE Commun. Surveys Tuts.}, vol.~22, no.~2, pp. 869--904, Jan. 2020.

\bibitem{Hardware_2020}
\BIBentryALTinterwordspacing
B.~Rutledge, ``New coral products for 2020,'' Google, Tech. Rep., jan. 2020.
  [Online]. Available:
  \url{https://developer.nvidia.com/blog/nvidia-jetson-agx-xavier-32-teraops-ai-robotics/}
\BIBentrySTDinterwordspacing

\bibitem{TensorFlow_Lite}
\BIBentryALTinterwordspacing
K.~LeViet, ``What's new in tensorflow lite from devsummit 2020,'' TensorFlow,
  Tech. Rep., Apr. 2020. [Online]. Available:
  \url{https://blog.tensorflow.org/2020/04/whats-new-in-tensorflow-lite-from-devsummit-2020.html}
\BIBentrySTDinterwordspacing

\end{thebibliography}

\end{document}